\documentclass[12pt,number,preprint,3p]{elsarticle}
\usepackage{titlesec}
\usepackage{color}
\usepackage[dvipsnames]{xcolor}
\usepackage{soul}
\usepackage{graphicx}
\usepackage{subcaption}
\usepackage[colorlinks]{hyperref}
\makeatletter
\gdef\urlauthor#1#2{\g@addto@macro\@elsuads{\let\corref\@gobble%
     \def\@@tmp{#1}\raggedright\eadsep
     {\ttfamily\url{\expandafter\strip@prefix\meaning\@@tmp}}\space(#2)%
     \def\eadsep{\unskip,\space}}%
}
\gdef\emailauthor#1#2{\stepcounter{ead}%
     \g@addto@macro\@elseads{\raggedright%
      \let\corref\@gobble\def\@@tmp{#1}%
      \eadsep{\ttfamily\href{mailto:\expandafter\strip@prefix\meaning\@@tmp}{\expandafter\strip@prefix\meaning\@@tmp}}
      (#2)\def\eadsep{\unskip,\space}}%
}
\makeatother
\usepackage{amssymb}
\usepackage{amsthm}
\usepackage{amsmath}
\usepackage{amssymb}
\usepackage{mathtools}
\usepackage{dsfont}
\usepackage{booktabs}
\usepackage{url}
\usepackage{scrextend}
\usepackage{epstopdf}
\usepackage{float}
\usepackage{tcolorbox}
\usepackage{tablefootnote}
\usepackage[perpage]{footmisc}
\usepackage{lineno}
\usepackage{multirow}
\usepackage{pifont}

\usepackage[ruled,vlined]{algorithm2e}

\newcommand{\vect}[1]{\boldsymbol{#1}}



\makeatletter
\let\@afterindenttrue\@afterindentfalse
\makeatother

\biboptions{numbers,sort&compress,square}

\usepackage{enumitem}

\setlist[itemize]{itemsep=3pt, topsep=3pt, parsep=3pt, partopsep=3pt}
\setlist[enumerate]{itemsep=3pt, topsep=3pt, parsep=3pt, partopsep=3pt}

\begin{document}

	\begin{frontmatter}
		\renewcommand{\thefootnote}{\fnsymbol{footnote}}
		\title{Surface decomposition method for sensitivity analysis of first-passage dynamic reliability of linear systems}
		\author[1]{Jianhua Xian}  
            \author[1]{Sai Hung Cheung\corref{cor1}}
            \ead{jhcheung@hku.hk}
            \cortext[cor1]{Corresponding author}
            \author[2,3,4]{Cheng Su}     
		\address[1]{Department of Civil Engineering, The University of Hong Kong, Hong Kong, China}
        \address[2]{State Key Laboratory of Subtropical Building and Urban Science, South China University of Technology, Guangzhou, China}
        \address[3]{School of Civil Engineering and Transportation, South China University of Technology, Guangzhou, China}
        \address[4]{School of Civil Engineering, Guangzhou City University of Technology, Guangzhou, China}

	\begin{abstract}
        This work presents a novel surface decomposition method for the sensitivity analysis of first-passage dynamic reliability of linear systems subjected to Gaussian random excitations. The method decomposes the sensitivity of first-passage failure probability into a sum of surface integrals over the constrained component limit-state hypersurfaces. The evaluation of these surface integrals can be accomplished, owing to the availability of closed-form linear expressions of both the component limit-state functions and their sensitivities for linear systems. An importance sampling strategy is introduced to further enhance the efficiency for estimating the sum of these surface integrals. The number of function evaluations required for the reliability sensitivity analysis is typically on the order of $10^2$ to $10^3$. The approach is particularly advantageous when a large number of design parameters are considered, as the results of function evaluations can be reused across different parameters. Three numerical examples are investigated to demonstrate the effectiveness of the proposed method.
        \end{abstract}
  
    
	\end{frontmatter}

	\renewcommand{\thefootnote}{\fnsymbol{footnote}}
	
\section{Introduction}
Reliability analysis is fundamental for assessing the safety of engineering structures due to the ubiquitous uncertainties arising in structural parameters such as material properties and damping ratios, as well as in external loadings such as wind, waves and ground excitations due to earthquakes \cite{ditlevsen1996structural,der1996structural}. In structural dynamics applications, the so-called first-passage dynamic reliability problems have been extensively investigated, where the system failure is defined as the event that any critical performance measure exceeds its prescribed threshold during the stochastic loading process \cite{li2009stochastic}. The first-passage dynamic reliability problems can be addressed by the classical level-crossing-process-based methods with certain assumptions \cite{chen1997study,kawano1999dynamic}, which, however, is generally limited to cases only involving a single response process. To overcome these limitations, various reliability approaches have been developed, such as the surrogate-modeling-based methods \cite{wang2017equivalent,xian2024physics,kim2024dimensionality}, probability-density-evolution-based methods \cite{chen2007extreme,chen2019direct,xian2021seismic,lyu2022unified} and sampling-based methods \cite{au2001estimation,bansal2017evaluation,xian2024relaxation}.

For first-passage dynamic reliability problems of linear systems subjected to Gaussian random excitations, several advanced sampling techniques have been developed to achieve extremely efficient and highly accurate estimates of the failure probability \cite{au2001first,misraji2020application,katafygiotis2006domain,valdebenito2021failure}. The common feature of these sampling techniques lies in that they fully exploit the system linearity and excitation Gaussianity. The efficient importance sampling \cite{au2001first} constructs its importance density as a weighted sum of the theoretically optimal importance densities associated with the component limit-state functions, in which the weight coefficients are proportional to the component failure probabilities. This idea was further extended in directional sampling, leading to the directional importance sampling \cite{misraji2020application}. The domain decomposition method \cite{katafygiotis2006domain} decomposes the first-passage failure probability into a sum of domain integrals that can be evaluated by the direct Monte Carlo simulation, and an importance sampling scheme is then introduced to estimate their total contribution. The multidomain line sampling \cite{valdebenito2021failure} also leverages the idea of domain decomposition, where the resulting domain integrals are solved by line sampling, and a similar importance sampling strategy is adopted to estimate
their sum.


Reliability sensitivity analysis aims to quantify and understand the influence of small perturbations of system parameters on the probability of system failure \cite{wu1994computational,lacaze2015probability,talebiyan2020sampling}, which is essential in risk-informed decision making \cite{song2009system,taflanidis2011simulation} and reliability-based design optimization \cite{youn2004selecting,chun2016structural}. Several sampling-based and surrogate-modeling-based methods have been developed to address the reliability sensitivity problems. In the former category, Lu et al. \cite{lu2008reliability} and Song et al. \cite{song2009subset} extended line sampling and subset simulation, respectively, for reliability sensitivity analysis with respect to the distribution parameters of basic random variables. Iason et al. \cite{papaioannou2018reliability} developed the sequential importance sampling for estimating reliability sensitivities with respect to the design parameters that appear in the indicator function, in which a smooth approximation of the indicator function is employed. In the latter category, Rahman \cite{rahman2009stochastic} investigated the moment and reliability sensitivities with respect to probability distribution parameters leveraging surrogate performance function constructed through dimensional decomposition. Dubourg and Sudret \cite{dubourg2014meta} incorporated the Kriging-based importance sampling with a score function approach to study the influence of the design parameters on the structural failure probability. Yun et al. \cite{yun2020adaptive} developed an approach combining adaptive subdomain sampling and adaptive Kriging construction to compute the failure probability sensitivities. Torii et al. \cite{torii2017probability} approximated both limit-state function and its derivative using polynomial expansions to facilitate sensitivity analysis of the structural failure probability. It is noted that, similar to their application in reliability analysis, the surrogate-modeling-based methods for reliability sensitivity analysis are generally restricted to problems with low-dimensional uncertain input.

Sensitivity analysis of the first-passage dynamic reliability, which represents a challenging branch of reliability sensitivity analysis, has received increasing attention in recent years \cite{au2005reliability,jensen2015reliability,chun2015parameter,yang2022structural,li2023direct,xian2023reliability}. In the context of sampling-based methods, Au \cite{au2005reliability} presented an efficient simulation approach designed for dynamic-reliability-based sensitivity analysis involving one or two parameters by artificially constructing an augmented reliability problem. Jensen et al. \cite{jensen2015reliability} investigated a subset-simulation-based approach to compute the sensitivities of the first-excursion failure probability with respect to the distribution parameters of uncertain design variables. Besides, Chun et al. \cite{chun2015parameter} developed a sequential compounding method for the parametric sensitivity analysis of general system problems, and then applied this method to a first-passage dynamic reliability problem of a linear system. Yang et al. \cite{yang2022structural} extended the probability density evolution method for evaluation of the first-passage dynamic reliability sensitivities of nonlinear structures with respect to deterministic design parameters. Li et al. \cite{li2023direct} also extended the direct probability integral method for first-passage failure probability sensitivity analysis of nonlinear structures. Xian and Su \cite{xian2023reliability} developed an efficient approach for sensitivity analysis of the first-excursion dynamic reliability of fractionally-damped structures incorporating  the explicit time-domain method, adjoint variable method and Poisson process method. Furthermore, building upon the developments confined to linear systems under Gaussian random excitations, the directional importance sampling and domain decomposition method have been extended from dynamic reliability analysis to reliability sensitivity analysis in \cite{valdebenito2021sensitivity} and \cite{misraji2025first}, respectively.

In essence, reliability sensitivity analysis entails the evaluation of a high-dimensional integral over the limit-state hypersurface, which is generally challenging to compute \cite{papaioannou2018reliability}, particularly for first-passage dynamic reliability problems characterized by highly non-smooth and nonlinear limit-state hypersurfaces. Fortunately, in the case of linear systems subjected to Gaussian excitations, the above complex surface integrals can be evaluated with remarkable efficiency. This work presents a novel surface decomposition method to accomplish this purpose. The core concept of this method is to decompose the complex surface integral into a sum of simpler surface integrals over the constrained component limit-state hypersurfaces. The so-called constrained component hypersurface corresponds exactly to the active part of the corresponding component hypersurface that contributes to the system limit-state hypersurface. The effectiveness of this strategy lies in the ease of computing these component surface integrals, as closed-form linear expressions of both the component limit-state functions and their sensitivities are available. An importance sampling scheme is further introduced to estimate the total contribution of these component surface integrals. The sensitivity analyses of the first-passage dynamic reliability problems for an oscillator, a shear-type structure and a building frame structure are conducted to demonstrate the feasibility of the proposed method.

\section{Explicit formulation of dynamic response and sensitivity for linear systems}
\subsection{Explicit expressions of dynamic responses}
The equation of motion for a linear system can be expressed as
\begin{equation}\label{motion equation}
    \vect M \ddot{\vect U}(\vect X ,t)+\vect C \dot{\vect U}(\vect X ,t)+\vect K \vect U(\vect X ,t)
    = \vect L F(\vect X ,t) 
\end{equation}
where $\vect M$, $\vect C$ and $\vect K$ are the mass, damping and stiffness matrices of the linear system, respectively; $\ddot{\vect U}(\vect X ,t)$, $\dot{\vect U}(\vect X ,t)$ and $\vect U(\vect X ,t)$ are the acceleration, velocity and displacement vectors of the linear system, respectively; $F(\vect X ,t)$ is the external excitation assumed to be a zero-mean Gaussian random process; $\vect X$ is the input random vector associated with the external excitation; and $\vect L$ is the orientation vector of $F(\vect X ,t)$.

Suppose that the linear system is initially at rest. Then, from the discrete perspective of the convolution integral, any response $r(\vect X ,t)$ of the linear system can be written in an explicit time-domain form as follows:
\begin{equation}\label{eq:2}
    r_i(\vect X)
    = a^r_{i,1}F_1(\vect X)+a^r_{i,2}F_2(\vect X) +\cdots+ a^r_{i,i}F_i(\vect X) \quad(i=1,2,\dots,n)
\end{equation}
where $n$ is the number of time steps for the time-history analysis; $r_i(\vect X)=r(\vect X,t_i)$ with $t_i=i\Delta t$ and $\Delta t$ being the time step; $F_j(\vect X)=F(\vect X,t_j)$ with $t_j=j\Delta t\;(j=1,2,\dots,i)$; $a^r_{i,j}\;(j=1,2,\dots,i)$ are the coefficients with respect to $r_i(\vect X)$.

The physical meaning and computational scheme of the coefficients $a^r_{i,j}\;(1\leq j\leq i\leq n)$ have been well discussed in \cite{su2014random}. For these coefficients, only $a^r_{i,1}\;(1\leq i\leq n)$ need to be calculated and stored, while the other coefficients can be determined from the recursive formula $a^r_{i,j}=a^r_{i-1,j-1}\;(2\leq j\leq i\leq n)$. The coefficients $a^r_{i,1}\;(1\leq i\leq n)$ represent the response $r(\vect X,t)$ of the linear system at time $t=t_1,t_2,\dots,t_n$ subjected to a unit impulse excitation applied at time $t=t_1$. Therefore, by performing a single response time-history analysis of the linear system shown in Eq.\eqref{motion equation} with any type of time-domain integration methods, the coefficients $a^r_{i,j}\;(1\leq j\leq i\leq n)$ with respect to any response $r(\vect X,t)$ of interest can be readily obtained.

The Gaussian random excitation $F(\vect X ,t)$ can be characterized by either its correlation function $R_F(t,\tau)$ or its power spectrum $S_F(\omega,t)$. Given the correlation function $R_F(t,\tau)$, the excitation at time $t=t_i$ can be represented, based on the orthogonal decomposition method \cite{grigoriu2006evaluation}, as
\begin{equation}\label{eq:3}
    F_i(\vect X)
    = \vect \psi_i \vect X \quad(i=1,2,\dots,n)
\end{equation}
where $\vect X$ is a standard Gaussian random vector with a dimension of $d=n$; and $\vect \psi_i$ is the $i$-th row vector of the matrix $\vect \Psi \vect\Lambda^{1/2}$, in which $\vect \Psi$ and $\vect\Lambda$ are respectively the orthonormal eigenvector matrix and the diagonal eigenvalue matrix of the following covariance matrix:
\begin{equation}
\vect \Sigma
= \begin{bmatrix}
R_F(t_1,t_1) & R_F(t_1,t_2) & \cdots & R_F(t_1,t_n) \\
R_F(t_2,t_1) & R_F(t_2,t_2) & \cdots & R_F(t_2,t_n) \\
\vdots & \vdots & \ddots & \vdots \\
R_F(t_n,t_1) & R_F(t_n,t_2) & \cdots & R_F(t_n,t_n)
\end{bmatrix}
\end{equation}

If the power spectrum $S_F(\omega,t)$ is available, the excitation at time $t=t_i$ can also be expressed in the same form as Eq.\eqref{eq:3} utilizing the spectral representation method \cite{deodatis2025spectral}. In this case, $\vect \psi_i$ is written as
\begin{equation}\label{eq:5}
\vect{\psi}_i =
\sqrt{2\Delta\omega}
\begin{bmatrix}
\sqrt{S_F(\omega_1,t_i)}\cos(\omega_1 t_i) \\
\sqrt{S_F(\omega_2,t_i)}\cos(\omega_2 t_i) \\
\vdots \\
\sqrt{S_F(\omega_q,t_i)}\cos(\omega_q t_i) \\
\sqrt{S_F(\omega_1,t_i)}\sin(\omega_1 t_i) \\
\sqrt{S_F(\omega_2,t_i)}\sin(\omega_2 t_i) \\
\vdots \\
\sqrt{S_F(\omega_q,t_i)}\sin(\omega_q t_i)
\end{bmatrix}^{\mathrm{T}} \quad (i=1,2,\dots,n)
\end{equation}
where $q$ is the number of circular frequency intervals; $\Delta \omega=(\omega_{\max}-\omega_{\min})/q$ is the circular frequency interval with $\omega_{\max}$ and $\omega_{\min}$ being the upper and lower cutoff circular frequencies, respectively; and $\omega_k=\omega_{\min}+(k-0.5)\Delta \omega \; (k=1,2,\dots,q)$ is the circular frequency in the middle interval. Note that, when $\vect \psi_i$ is expressed in Eq.\eqref{eq:5}, the dimension of the standard Gaussian random vector $\vect X$ is $d=2q$.

Substitution of Eq.\eqref{eq:3} into Eq.\eqref{eq:2} yields a more compact explicit expression of $r(\vect X,t)$ as follows:
\begin{equation}\label{explicit expression}
    r_i(\vect X)
    = \vect a_i^r \vect X \quad(i=1,2,\dots,n)
\end{equation}
where
\begin{equation}\label{coefficient vector a}
    \vect a_i^r=a^r_{i,1}\vect \psi_1+a^r_{i,2}\vect \psi_2 +\cdots+ a^r_{i,i}\vect \psi_i
\end{equation}
is the coefficient vector with respect to $r_i(\vect X)$. Eq.\eqref{explicit expression} indicates that the response of the linear system at any given time instant can be written as a linear combination of all the random variables contained in the input random vector $\vect X$.

It is noted that, for brevity, this study merely focuses on a single random excitation in the problem formulation. However, the extension to cases involving multiple random excitations is straightforward, with the number of response time-history analyses required for explicit formulation being equal to the number of random excitations.

\subsection{Explicit expressions of response sensitivities}
Suppose $\theta$ is a design parameter of the linear system shown in Eq.\eqref{motion equation}. Then, differentiating all the terms in Eq.\eqref{motion equation} with respect to the design parameter $\theta$, one can derive the sensitivity equation for the linear system as follows:
\begin{equation}\label{sensitivity equation}
\begin{aligned}
    &\vect M\frac{\partial\ddot{\vect U}(\vect X ,t)}{\partial \theta} 
    +\vect C\frac{\partial\dot{\vect U}(\vect X ,t)}{\partial \theta} 
    + \vect K\frac{\partial\vect U(\vect X ,t)}{\partial \theta} \\
    &= \frac{\partial\vect L}{\partial \theta} F(\vect X ,t) 
    -\left\{ \frac{\partial\vect M}{\partial \theta} \ddot{\vect U}(\vect X ,t)
    +\frac{\partial\vect C}{\partial \theta} \dot{\vect U}(\vect X ,t)
    + \frac{\partial\vect K}{\partial \theta} \vect U(\vect X ,t) \right\}
\end{aligned}
\end{equation}
where $\partial\vect M/\partial \theta$, $\partial\vect C/\partial \theta$, $\partial\vect K/\partial\theta$ and $\partial\vect L/\partial \theta$ denote the sensitivities of the matrices $\vect M$, $\vect C$, $\vect K$ and $\vect L$ with respect to $\theta$, respectively; and $\partial\ddot{\vect U}(\vect X ,t)/\partial \theta$, $\partial\dot{\vect U}(\vect X ,t)/\partial \theta$ and $\partial\vect U(\vect X ,t)/\partial \theta$ denote the sensitivities of the response vectors $\ddot{\vect U}(\vect X ,t)$, $\dot{\vect U}(\vect X ,t)$ and $\vect U(\vect X ,t)$ with respect to $\theta$, respectively.

Differentiation of both sides of Eq.\eqref{eq:2} with respect to the design parameter $\theta$ yields the explicit time-domain form of the response sensitivity $\partial r(\vect X,t)/\partial \theta$ as follows:
\begin{equation}\label{explicit response sensitivity}
    \frac{\partial r_i(\vect X)}{\partial \theta}
    = b^r_{i,1}F_1(\vect X)+b^r_{i,2}F_2(\vect X) +\cdots+ b^r_{i,i}F_i(\vect X) \quad(i=1,2,\dots,n)
\end{equation}
where $b^r_{i,j}=\partial a^r_{i,j}/\partial \theta\;(j=1,2,\dots,i)$ are the coefficients with respect to $\partial r_i(\vect X)/\partial \theta$.

The coefficients $b^r_{i,j}\;(1\leq j\leq i\leq n)$ can be derived via direct differentiation of the coefficients $a^r_{i,j}\;(1\leq j\leq i\leq n)$ in Eq.\eqref{eq:2} with respect to $\theta$. However, this requires knowledge of the analytical formulas of $a^r_{i,j}\;(1\leq j\leq i\leq n)$. In fact, similar to the coefficients $a^r_{i,j}\;(1\leq j\leq i\leq n)$, the coefficients $b^r_{i,j}\;(1\leq j\leq i\leq n)$ also possess inherent physical meanings, and a detailed computational scheme of these coefficients have been presented in \cite{hu2016explicit}. Among these coefficients, only $b^r_{i,1}\;(1\leq i\leq n)$ need to be calculated and stored, while a recursive formula $b^r_{i,j}=b^r_{i-1,j-1}\;(2\leq j\leq i\leq n)$ can be adopted to determine the other coefficients. The coefficients $b^r_{i,1}\;(1\leq i\leq n)$ represent the response sensitivity $\partial r(\vect X,t)/\partial \theta$ of the linear system at time $t=t_1,t_2,\dots,t_n$ driven by a unit impulse excitation applied at time $t=t_1$. Therefore, by conducting a single response sensitivity time-history analysis based on the sensitivity equation of the linear system presented in Eq.\eqref{sensitivity equation}, the coefficients $b^r_{i,j}\;(1\leq j\leq i\leq n)$ can be easily obtained for the sensitivity of any response $\partial r(\vect X,t)/\partial \theta$. The above response sensitivity time-history analysis can be achieved using any type of time-domain integration methods.

Similar to the derivation of Eq.\eqref{explicit expression}, substitution of Eq.\eqref{eq:3} into Eq.\eqref{explicit response sensitivity} results in  the following compact explicit expression of $\partial r(\vect X,t)/\partial \theta$ as
\begin{equation}\label{explicit sensitivity}
    \frac{\partial r_i(\vect X)}{\partial \theta}
    = \vect b_i^r \vect X \quad(i=1,2,\dots,n)
\end{equation}
where
\begin{equation}\label{coefficient vector partial a}
    \vect b_i^r= b^r_{i,1}\vect \psi_1+ b^r_{i,2}\vect \psi_2 +\cdots+  b^r_{i,i}\vect \psi_i
\end{equation}
is the coefficient vector with respect to $\partial r_i(\vect X)/\partial \theta$. 

It is noted that, when $\theta$ is associated with the external excitation $F(\vect X ,t)$, for instance as a parameter defining the correlation function $R_F(t,\tau)$ or the power spectrum $S_F(\omega,t)$, the explicit expression of $\partial r(\vect X,t)/\partial \theta$ can be obtained by direct differentiation of Eqs.\eqref{explicit expression} and \eqref{coefficient vector a}, without performing any additional response sensitivity time-history analysis of the linear system.

Thus far, the explicit expressions for both the dynamic responses and their sensitivities have been constructed for the linear system, as shown in Eqs.\eqref{explicit expression} and \eqref{explicit sensitivity}, respectively. The availability of such concise closed-form expressions greatly facilitates the sensitivity analysis of first-passage dynamic reliability, which will be demonstrated in Section \ref{sec:surface decomposition}.

\section[Surface decomposition method for sensitivity analysis of first-passage dynamic reliability]%
{Surface decomposition method for sensitivity analysis of first-passage \\ dynamic reliability}\label{sec:surface decomposition}
\subsection{First-passage dynamic reliability}
For first-passage dynamic reliability problems of linear systems, the system failure event is typically defined as the exceedance of the prescribed threshold by any response component of interest at any time instant within a specified time duration. The system failure event can be formulated as the union of the component failure events, i.e.
\begin{equation}\label{system failure event}
    F=\bigcup_{k=1}^{m}\bigcup_{i=1}^{n}F_{ki}
\end{equation}
where $m$ is the number of critical response components; $n$ is the number of time steps for the given time interval; and $F_{ki}$ is the component failure event corresponding to the $k$th critical response component $s_k(\vect X,t)$ at the time instant $t_i$, which can be defined as
\begin{equation}\label{component failure event}
    F_{ki}=\left\{ c_k -  s_{k,i}(\vect X)  \leq0 \right\}\quad(i=1,2,\dots,n;k=1,2,\dots,m)
\end{equation}
where $s_{k,i}(\vect X)=s_{k}(\vect X,t_i)$; and $c_k$ is the positive threshold for the $k$th critical response component. Note that the formulation is presented for up-crossing of positive thresholds, while extension to down-crossing of negative thresholds or simultaneous consideration of both cases is straightforward.

The probability of the component failure event $F_{ki}$ is given by
\begin{equation}
    \mathbb{P}_{ki}
    = \int_{\mathbb{R}^n} \mathcal{H}(-g_{ki}(\vect x)) f_{\vect X}(\vect x)\, d\vect x\quad(i=1,2,\dots,n;k=1,2,\dots,m)
\end{equation}
where 
\begin{equation}\label{component lsf}
    g_{ki}(\vect x) = c_k - s_{k,i}(\vect x)\quad(i=1,2,\dots,n;k=1,2,\dots,m)
\end{equation}
is the component limit-state function with respect to $F_{ki}$; $f_{\vect X}(\vect x)$ is the joint probability density function (PDF) of the standard Gaussian random vector $\vect X$; and $\mathcal{H}(\cdot)$ denotes the Heaviside step function, with $\mathcal{H}(-g_{ki}(\vect x))$  taking 1 if $g_{ki}(\vect x)\leq 0$ and 0 otherwise.

Let $r(\vect X,t)$ be $s_k(\vect X,t)$. Substituting Eq.\eqref{explicit expression} into Eq.\eqref{component lsf}, one has
\begin{equation}\label{component lsf1}
    g_{ki}(\vect x) = c_k - \vect a_i^{s_k} \vect x\quad(i=1,2,\dots,n;k=1,2,\dots,m)
\end{equation}
Since the limit-state hypersurface described by $g_{ki}(\vect x) = c_k - \vect a_i^{s_k} \vect x=0$ is a  hyperplane, closed-form solution to the failure probability $\mathbb{P}_{ki}$ can be obtained as
\begin{equation}\label{component probability}
    \mathbb{P}_{ki} = \Phi (-\beta_{ki})\quad(i=1,2,\dots,n;k=1,2,\dots,m)
\end{equation}
where $\Phi (\cdot)$ is the standard Gaussian cumulative distribution function (CDF); and $\beta_{ki}$ is the reliability index with respect to $F_{ki}$, which can be expressed as
\begin{equation}\label{beta}
    \beta_{ki} = \frac {\mu_{g_{ki}}}{\sigma _{g_{ki}}}= \frac{c_k}{\left\| \vect a_i^{s_k} \right\|}\quad(i=1,2,\dots,n;k=1,2,\dots,m)
\end{equation}
where $\mu_{g_{ki}}$ and $\sigma _{g_{ki}}$ are the mean and standard deviation of the Gaussian random variable $g_{ki}(\vect X)$; and $\left\| \cdot \right\|$ is the Euclidean norm.

The reliability index $\beta_{ki}$ can be interpreted geometrically as the Euclidean distance between the origin and the design point $\vect{x}_{ki}^{\star}$. The design point $\vect{x}_{ki}^{\star}$, also referred to as the most probable point, is defined as the point on the limit-state hypersurface $g_{ki}(\vect{x}) = 0$ that lies closest to the origin and can be expressed as
\begin{equation}\label{design point}
    \vect{x}_{ki}^{\star} =\beta_{ki}\vect{u}_{ki} \quad(i=1,2,\dots,n;k=1,2,\dots,m)
\end{equation}
where $\vect{u}_{ki}$ is the unit vector pointing from the origin to the design point, which is aligned with the normalized gradient of the limit-state function $g_{ki}(\vect{x})$ and hence perpendicular to the hyperplane $g_{ki}(\vect{x}) = 0$. The unit vector is given by
\begin{equation}\label{unit vector}
    \vect{u}_{ki} =- \frac {\nabla _{\vect x}g_{ki}(\vect{x})}{\left\|\nabla _{\vect x}g_{ki}(\vect{x})\right\|}= \frac{\vect a_i^{s_k}}{\left\| \vect a_i^{s_k} \right\|}\quad(i=1,2,\dots,n;k=1,2,\dots,m)
\end{equation}
where $\nabla _{\vect x}g_{ki}(\vect{x})$ denotes the gradient of $g_{ki}(\vect{x})$.

Although the probabilities of individual component failure events can be readily computed in closed form by Eqs.\eqref{component probability} and \eqref{beta}, an analytical evaluation of the system failure probability is not feasible. The probability of the system failure event $F$ can be expressed as
\begin{equation}\label{system failure probability}
    \mathbb{P}
    = \int_{\mathbb{R}^n} \mathcal{H}(-G(\vect x)) f_{\vect X}(\vect x)\, d\vect x
\end{equation}
where $G(\vect x)$ is the system limit-state function written as
\begin{equation}\label{system lsf}
    G(\vect x)= \min_{k=1}^{m}\min_{i=1}^{n}g_{ki}(\vect x)
\end{equation}

By exploiting the system linearity and leveraging the analytical solutions to the component failure probabilities, certain advanced simulation techniques, such as the efficient importance sampling \cite{au2001first} and the domain decomposition method \cite{katafygiotis2006domain}, were proposed for solving the probability integral in Eq.\eqref{system failure probability} at high efficiency.

\subsection{Sensitivity of first-passage dynamic reliability}
Differentiation of the system failure probability formulated in Eq.\eqref{system failure probability} with respect to the design parameter $\theta$ leads to the sensitivity of system failure probability as follows:
\begin{equation}\label{system failure probability sensitivity}
    \frac{\partial\mathbb{P}}{\partial \theta}
    = -\int_{\mathbb{R}^n} \delta(-G(\vect x)) \frac{\partial G(\vect x)}{\partial \theta}f_{\vect X}(\vect x)\, d\vect x
\end{equation}
where $\delta(\cdot)$ denotes the Dirac delta function; and $\partial G(\vect x)/\partial \theta$ is the sensitivity of the system limit-state function $G(\vect x)$ with respect to $\theta$.

Introduce an indicator function $\mathbb{I}_{ki}(\vect x)$, which takes 1 if the corresponding component limit-state function $g_{ki}(\vect x)$ attains the minimum among all component limit-state function, and 0 otherwise. The indicator function can be mathematically expressed as
\begin{equation}\label{indicator}
\mathbb{I}_{ki}(\vect x) =
\begin{cases}
1, & \text{if } g_{ki}(\vect x) = \min\limits_{j=1}^{m}\min\limits_{s=1}^{n} g_{js}(\vect x), \\[6pt]
0, & \text{otherwise}.
\end{cases}\quad(i=1,2,\dots,n;k=1,2,\dots,m)
\end{equation}
The above indicator functions serve to partition the integration domain of Eq.\eqref{system failure probability sensitivity} into $mn$ mutually exclusive and collectively exhaustive subdomains $\Omega_{ki}= \left\{ \vect x\in  \mathbb{R}^n: \mathbb{I}_{ki}(\vect x) =1\right\}$ $(i=1,2,\dots,n;k=1,2,\dots,m)$, within each of which only the component attaining the minimum remains active.

With the indicator functions shown in Eq.\eqref{indicator}, the sensitivity of system failure probability shown in Eq.\eqref{system failure probability sensitivity} can be expanded as
\begin{equation}\label{system failure probability sensitivity expanded}
    \frac{\partial\mathbb{P}}{\partial \theta}
    = \sum_{k=1}^{m}\sum_{i=1}^{n}\eta_{ki} 
\end{equation}
where
\begin{equation}\label{eta0}
    \eta_{ki}
    =  -\int_{\mathbb{R}^n} \delta(-G(\vect x)) \frac{\partial G(\vect x)}{\partial \theta}\mathbb{I}_{ki}(\vect x)f_{\vect X}(\vect x)\, d\vect x \quad(i=1,2,\dots,n;k=1,2,\dots,m)
\end{equation}
denotes the $d$-dimensional volume integral over the subdomain $\Omega_{ki}$.

Within each subdomain $\Omega_{ki}$, one has $G(\vect x)=g_{ki}(\vect x)$, and hence Eq.\eqref{eta0} can be further written as
\begin{equation}\label{eta}
    \eta_{ki}
    =  -\int_{\mathbb{R}^n} \delta(-g_{ki}(\vect x)) \frac{\partial g_{ki}(\vect x)}{\partial \theta}\mathbb{I}_{ki}(\vect x)f_{\vect X}(\vect x)\, d\vect x \quad(i=1,2,\dots,n;k=1,2,\dots,m)
\end{equation}
where $\partial g_{ki}(\vect x)/\partial \theta$ is the sensitivity of the component limit-state function $g_{ki}(\vect x)$ with respect to $\theta$, which can be derived from Eq.\eqref{component lsf} as follows:
\begin{equation}\label{sensitivity of component lsf}
    \frac{\partial g_{ki}(\vect x)}{\partial \theta} = - \frac{\partial s_{k,i}(\vect x)}{\partial \theta}\quad(i=1,2,\dots,n;k=1,2,\dots,m)
\end{equation}
where $\partial s_{k,i}(\vect x)/\partial \theta$ is the sensitivity of the $k$th critical response component $s_k(\vect X,t)$ at the time instant $t_i$ with respect to $\theta$. Let $\partial r(\vect X,t)/\partial \theta = \partial s_k(\vect X,t)/\partial \theta$. Substituting Eq.\eqref{explicit sensitivity} into Eq.\eqref{sensitivity of component lsf}, one gets
\begin{equation}\label{sensitivity of component lsf 1}
    \frac{\partial g_{ki}(\vect x)}{\partial \theta} = - \vect b_i^{s_k} \vect x \quad(i=1,2,\dots,n;k=1,2,\dots,m)
\end{equation}

Since each component limit-state function $g_{ki}(\vect x)$ is continuously differentiable with a non-zero constant gradient, the $d$-dimensional volume integral in Eq.\eqref{eta} can be converted into a $(d-1)$-dimensional surface integral as follows \cite{hormander2003analysis}:
\begin{equation}\label{surface integral}
    \eta_{ki}
    = -\int_{g_{ki}(\vect x)=0} \frac{1}{\left\|\nabla _{\vect x}g_{ki}(\vect x)\right\|} \frac{\partial g_{ki}(\vect x)}{\partial \theta}\mathbb{I}_{ki}(\vect x)f_{\vect X}(\vect x)\, d S(\vect x) \quad(i=1,2,\dots,n;k=1,2,\dots,m)
\end{equation}
where $d S(\vect x)$ denotes the differential element of the $(d-1)$-dimensional hyperplane defined by $g_{ki}(\vect x)=0$; and $\left\|\nabla _{\vect x}g_{ki}(\vect x)\right\|$ is the Euclidean norm of the gradient of the component limit-state function $g_{ki}(\vect x)$, which can be obtained from Eq.\eqref{component lsf1} as follows:
\begin{equation}\label{gradient}
    \left\|\nabla _{\vect x}g_{ki}(\vect x)\right\|
    = \left\|\vect a_i^{s_k}\right\|\quad(i=1,2,\dots,n;k=1,2,\dots,m)
\end{equation}

Since the integration in Eq.\eqref{surface integral} is performed over the $(d-1)$-dimensional hyperplane $g_{ki}(\vect x)=0$, the indicator function shown in Eq.\eqref{indicator} can be further expressed as
\begin{equation}\label{indicator1}
\begin{aligned}
\mathbb{I}_{ki}(\vect x) &=
\begin{cases}
1, & \text{if }  g_{js}(\vect x)>0,\;\forall (j,s)\neq(k,i), \\[6pt]
0, & \text{otherwise}.
\end{cases} \quad (i=1,2,\dots,n; k=1,2,\dots,m)
\end{aligned}
\end{equation}
Essentially, the indicator function $\mathbb{I}_{ki}(\vect x)$ imposes additional constraints, restricting the integration to the portion of the hyperplane where all  other component limit-state functions remain in their safe domains. Such constrained hyperplanes are denoted as $S_{ki}=\left \{ \vect x \in \mathbb{R}^n:g_{ki}(\vect x)\mathbb{I}_{ki}(\vect x)=0\right\}\;(i=1,2,\dots,n; k=1,2,\dots,m)$.

\begin{figure}[ht]
  \centering
  \captionsetup{
  }
  \includegraphics[width=0.8\textwidth] {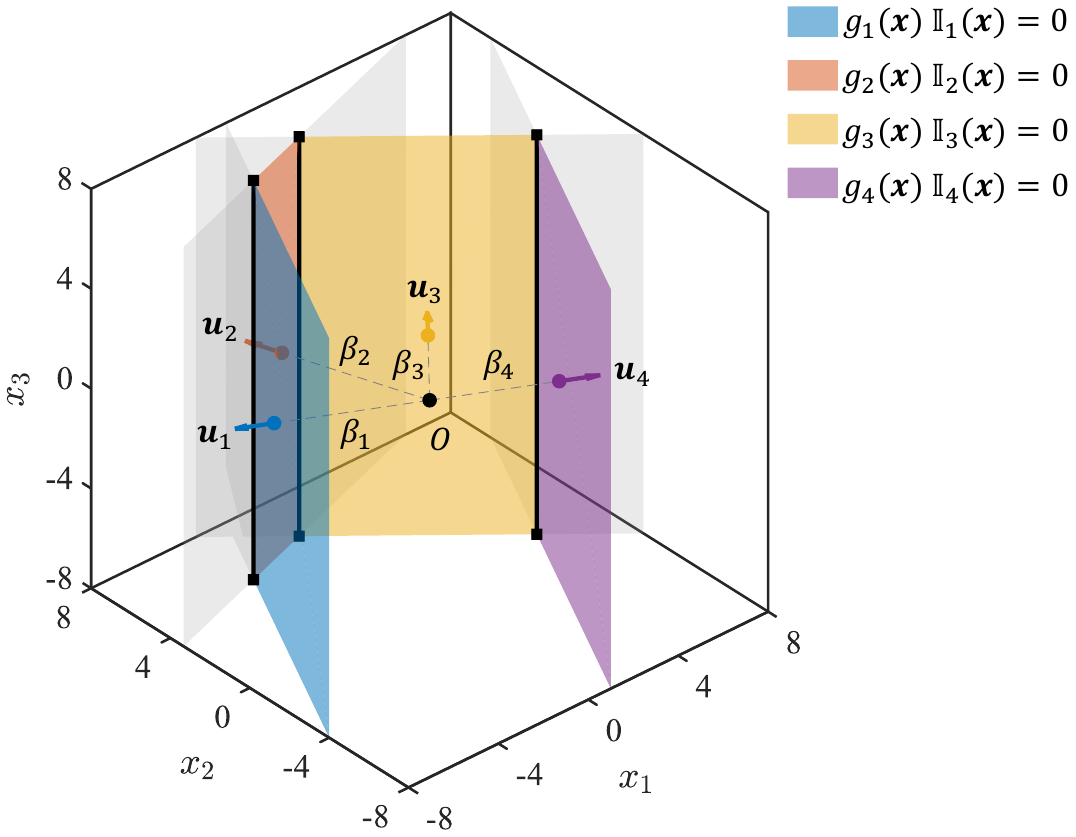}
  \caption{\textbf{Illustration of surface decomposition method using a 3-dimensional problem.} Consider a series system limit-state function $G(\vect x)=\min_{k=1}^{4} g_k(\vect x)$. The four component limit-state functions are $g_1(\vect x)=12-(-2x_1+x_2+0x_3)$, $g_2(\vect x)=18-(-x_1+3x_2+0x_3)$, $g_3(\vect x)=36-(6x_1+7x_2+0x_3)$ and $g_4(\vect x)=10-(2x_1-x_2+0x_3)$, respectively. The system failure surface is composed of four constrained component failure planes $g_k(\vect x)\mathbb{I}_k(\vect x)=0\;(k=1,2,3,4)$, in which the indicator function $\mathbb{I}_k(\vect x)$ takes 1 if $g_j(\vect x)>0, \forall j\neq k$, and 0 otherwise. The sensitivity of series system failure probability can be decomposed into four surface integrals over these four constrained component failure planes, respectively. These component failure planes will become component failure hyperplanes in higher-dimensional problems. }
  \label{fig:1}
\end{figure}

Note that the constrained hyperplane $S_{ki}$ corresponds exactly to the portion of the system limit-state hypersurface contributed by the corresponding component limit-state hypersurface. The union of all constrained hyperplanes $S_{ki}\;(i=1,2,\dots,n; k=1,2,\dots,m)$, which are mutually exclusive, constitutes the complete system limit-state hypersurface. 
Consequently, computing the sensitivity of the first-passage dynamic reliability amounts to integrating over a non-smooth system limit-state hypersurface. In this sense, Eqs.\eqref{system failure probability sensitivity expanded} and \eqref{surface integral} can be interpreted as decomposing this complex surface integral into a set of simpler integrals over the constrained hyperplanes—a strategy referred to as the \textit{surface decomposition method}. The basic idea of the proposed surface decomposition method is further illustrated in Figure \ref{fig:1} using a 3-dimensional toy problem.

By substituting Eqs.\eqref{sensitivity of component lsf 1} and \eqref{gradient} into Eq.\eqref{surface integral}, and applying the law of conditional probability, the surface integral $\eta_{ki}$ can be further expressed as
\begin{equation}\label{surface integral_sub}
\begin{split}
\eta_{ki}
= -\int_{g_{ki}(\vect x)=0}
\frac{\vect b_i^{s_k} \vect x}{\left\|\vect a_i^{s_k}\right\|}
\mathbb{I}_{ki}(\vect x)
f_{\vect X}(\vect x|\vect x^{\mathrm{T}}\vect{u}_{ki}=\beta_{ki})
\varphi (\beta_{ki})\, d S(\vect x)\\
\hfill (i=1,2,\dots,n;\;k=1,2,\dots,m)
\end{split}
\end{equation}
where $\varphi(\cdot)$ is the standard Gaussian PDF; $\beta_{ki}$ and $\vect{u}_{ki}$ are the reliability index and the unit vector associated with the component limit-state function $g_{ki}(\vect x)$, whose closed-form solutions are given in Eqs.\eqref{beta} and \eqref{unit vector}, respectively; and $f_{\vect X}(\vect x|\vect x^{\mathrm{T}}\vect{u}_{ki}=\beta_{ki})$ is the PDF of $\vect X$ conditional on the fact that it lies on the hyperplane $g_{ki}(\vect x)=0$.

Thus far, the surface integral $\eta_{ki}$ in Eq.\eqref{surface integral_sub} can be directly estimated with the Monte Carlo simulation as follows:
\begin{equation}\label{surface integral MCS}
    \eta_{ki}
    \approx \frac{1}{N_{ki}}\sum_{j=1}^{N_{ki}} \frac{\vect b_i^{s_k} \vect x_j}{\left\|\vect a_i^{s_k}\right\|} \mathbb{I}_{ki}(\vect x_j) \varphi (\beta_{ki})\quad(i=1,2,\dots,n;k=1,2,\dots,m)
\end{equation}
where $\vect x_j\;(j=1,2,\dots,N_{ki})$ denote samples generated from the PDF $f_{\vect X}(\vect x|\vect x^{\mathrm{T}}\vect{u}_{ki}=\beta_{ki})$, with $N_{ki}$ being the number of samples.

Simulating sample points from $f_{\vect X}(\vect x|\vect x^{\mathrm{T}}\vect{u}_{ki}=\beta_{ki})$ is equivalent to generating sample points from $f_{\vect X}(\vect x)$ and then projecting them onto the hyperplane $g_{ki}(\vect x)=0$, which can be accomplished as follows: i) generate a sample point $\vect x$ based on the PDF $f_{\vect X}(\vect x)$; ii) 
decompose $\vect x$ into a component that aligns with the unit vector $\vect{u}_{ki}$, $\vect x_{\parallel }=(\vect x^{\mathrm{T}}\vect{u}_{ki})\vect{u}_{ki}$, and a component that is perpendicular to  $\vect x_{\parallel}$, $\vect x_{\perp}=\vect x-\vect x_{\parallel}$; iii) construct ${\vect x}'=\beta_{ki}\vect{u}_{ki}+\vect x_{\perp}$, which represents a sample point  independently generated from the conditional PDF $f_{\vect X}(\vect x|\vect x^{\mathrm{T}}\vect{u}_{ki}=\beta_{ki})$.

It should be noted that the computation of the surface integral $\eta_{ki}$ via Monte Carlo simulation is trivial, as generating samples that fall into the mutual safe domain of all other component limit-state functions poses no particular difficulty. However, if the number of the surface integrals $\eta_{ki}\;(i=1,2,\dots,n;k=1,2,\dots,m)$ is large, the computational cost for the sensitivity of system failure probability shown in Eq.\eqref{system failure probability sensitivity expanded} could still be significant. To reduce the computational effort, an importance sampling strategy that has been developed in the domain decomposition method \cite{katafygiotis2006domain} for first-passage dynamic reliability analysis is introduced herein. The sum in Eq.\eqref{system failure probability sensitivity expanded} can be rewritten as
\begin{equation}\label{system failure probability sensitivity expanded rewritten}
    \frac{\partial\mathbb{P}}{\partial \theta}
    = mn\sum_{k=1}^{m}\sum_{i=1}^{n}\eta_{ki} f(k,i)
\end{equation}
where $f(k,i)=1/(mn)$ defines a discrete uniform probability mass function (PMF) over the index pairs $(k,i)\;(i=1,2,\dots,n;k=1,2,\dots,m)$.

Introducing an importance sampling PMF $h(k,i)$ to Eq.\eqref{system failure probability sensitivity expanded rewritten}, one yields 
\begin{equation}\label{system failure probability sensitivity IS}
    \frac{\partial\mathbb{P}}{\partial \theta}
    = mn\sum_{k=1}^{m}\sum_{i=1}^{n}\eta_{ki} \frac{f(k,i)}{h(k,i)}h(k,i)=\sum_{k=1}^{m}\sum_{i=1}^{n}\eta_{ki} \frac{1}{h(k,i)}h(k,i)\approx \frac{1}{N}\sum_{j=1}^{N}\eta_{K_j I_j}\frac{1}{h(K_j,I_j)}
\end{equation}
where $(K_j,I_j)\;(j=1,2,\dots,N)$ are the index pairs sampled from $h(k,i)$, with $N$ being the number of samples; and the surface integral $\eta_{K_j I_j}$ can be estimated by Eq.\eqref{surface integral MCS}.

There exists a theoretically optimal importance sampling PMF that results in a zero-variance estimate, i.e.
\begin{equation}\label{optimal ID}
    h^{\star}(k,i)
    = \frac{\eta_{ki}}{ \sum_{j=1}^{m}\sum_{s=1}^{n}\eta_{js} }
\end{equation}
However, direct use of this optimal importance sampling PMF is not possible, since the denominator $\sum_{j=1}^{m}\sum_{s=1}^{n}\eta_{js}$ is exactly the sensitivity of system failure probability $\partial\mathbb{P}/\partial \theta$ to be estimated.

A good choice of the importance sampling PMF is to construct it based on the closed-form solutions of the component failure probabilities, i.e. 
\begin{equation}\label{constructed ID}
    h(k,i)
    = \frac{\mathbb{P}_{ki}}{ \sum_{j=1}^{m}\sum_{s=1}^{n}\mathbb{P}_{js} }
\end{equation}
The rationale of this importance sampling PMF lies in the fact that a higher component failure probability $\mathbb{P}_{ki}$ typically indicates a larger measure of the constrained hyperplane $S_{ki}$. Consequently, the corresponding surface integral $\eta_{ki}$ contributes more significantly to the sensitivity of system failure probability.

Finally, the sampling scheme described by Eqs.\eqref{surface integral MCS} and \eqref{system failure probability sensitivity IS} can be combined into a single Monte Carlo estimator as follows:
\begin{equation}\label{final estimator}
    \frac{\partial\mathbb{P}}{\partial \theta}
    \approx \frac{1}{N}\sum_{j=1}^{N}\frac{1}{h(K_j,I_j)}
    \frac{\vect b_{I_j}^{s_{K_j}} \vect x_j}{\left\|\vect a_{I_j}^{s_{K_j}}\right\|} \mathbb{I}_{K_j I_j}(\vect x_j) \varphi (\beta_{K_j I_j})
\end{equation}
where $N$ is the number of samples; $(K_j,I_j)$ is the index pair sampled from the importance sampling PMF $h(k,i)$ in Eq.\eqref{constructed ID}; and $\vect x_j$ is sampled from the conditional PDF $f_{\vect X}(\vect x|\vect x^{\mathrm{T}}\vect{u}_{K_jI_j}=\beta_{K_jI_j})$.

It should be noted that evaluating the indicator function $\mathbb{I}_{ki}(\vect x)$ in Eq.\eqref{indicator1} is equivalent to calling the system limit-state function $G(\vect x)$ in Eq.\eqref{system lsf}, as both operations necessitate the computation of all component limit-state functions. Apart from evaluating the indicator function $\mathbb{I}_{ki}(\vect x)$, the computational cost of the other operations in Eq.\eqref{final estimator} is negligible. Therefore, the overall computational cost of the estimator in Eq.\eqref{final estimator} amounts to $N$ evaluations of the system limit-state function $G(\vect x)$. 

Furthermore, it can be seen from Eq.\eqref{final estimator} that the results of the indicator functions can be reused across different design parameters, meaning that the computational cost does not increase as the number of design parameters increases. This feature is particularly advantageous when sensitivities with respect to a large number of design parameters are required, as is commonly the case in reliability-based topology optimization.

For clarity, a computational procedure of the proposed surface decomposition method for the sensitivity analysis of first-passage dynamic reliability of linear systems is given as follows:

i) Perform a single impulse response time-history analysis of the linear system, and obtain the coefficients $a^{s_k}_{i,j}\;(1\leq j\leq i\leq n)$ for all the responses of interest $s_k(\vect X,t)\;(k=1,2,\dots,m)$. Construct $\vect a_i^{s_k}\;(i=1,2,\dots,n;k=1,2,\dots,m)$ by Eq.\eqref{coefficient vector a}.

ii) For each design parameter $\theta$, perform a single impulse response sensitivity time-history analysis and obtain the coefficients $b^{s_k}_{i,j}\;(1\leq j\leq i\leq n)$ for the response sensitivities $\partial s_k(\vect X,t)/\partial \theta \;(k=1,2,\dots,m)$. Construct $\vect b_i^{s_k}\;(i=1,2,\dots,n;k=1,2,\dots,m)$ by Eq.\eqref{coefficient vector partial a}.

iii) Compute the failure probabilities $\mathbb{P}_{ki}$, reliability indexes $\beta_{ki}$ and unit vectors $\vect u_{ki}$ by Eqs.\eqref{component probability}, \eqref{beta} and \eqref{unit vector}, respectively, for all component limit-state functions $g_{ki}(\vect x)\;(i=1,2,\dots,n;k=1,2,\dots,m)$.

iv) Construct the importance sampling PMF $h(k,i)$ by Eq.\eqref{constructed ID}.

v) Perform Algorithm \ref{algorithm} and obtain the estimate of the sensitivity of system failure probability $\partial\mathbb{P}/\partial \theta$ for each design parameter $\theta$.

\begin{algorithm}[htbp]
\caption{Proposed Algorithm for Surface Decomposition Method} 
\label{algorithm}
\KwIn{$\vect a_i^{s_k}$, $\vect b_i^{s_k}$ for each $\theta$, $\mathbb{P}_{ki}$, $\beta_{ki}$, $\vect u_{ki}$, $h(k,i)$, $tol=0.1$, $N_{\max}=10^4$}
\KwOut{$\partial\mathbb{P}/\partial \theta$ for each $\theta$, $N$}
\BlankLine
\For{$j \leftarrow 1$ \KwTo $N_{\max}$}{
    Generate index pair $(K_j,I_j) \sim h(k,i)$\;
    Generate sample $\vect x_j \sim f_{\vect X} (\vect x)$ and compute:\\
    \makebox[\linewidth][c]{
    $\vect x_j \leftarrow 
    \beta_{K_j I_j}\vect{u}_{K_j I_j} +
    \big(\vect x_j - (\vect x_j^{\mathrm{T}}\vect{u}_{K_j I_j})\vect{u}_{K_j I_j}\big);$
    }\\
    Compute indicator function $\mathbb{I}_{K_j I_j}(\vect x_j)$\;
    For each $\theta$, compute:\\
    \makebox[\linewidth][c]{
    $\mathrm{temp}_j[\theta] \leftarrow 
    \dfrac{1}{h(K_j,I_j)}
    \dfrac{\vect b_{I_j}^{s_{K_j}} \vect x_j}{\left\|\vect a_{I_j}^{s_{K_j}}\right\|} \mathbb{I}_{K_j I_j}(\vect x_j) \varphi (\beta_{K_j I_j}),$
    }\\
    
    \makebox[\linewidth][c]{
    $\mu_j[\theta] \leftarrow 
    \dfrac{1}{j}\sum_{s=1}^j\mathrm{temp}_s[\theta],$
    }\\

    \makebox[\linewidth][c]{
    $\delta_j[\theta] \leftarrow \dfrac{\sqrt{\dfrac{1}{j-1} \sum_{s=1}^{j} (\mathrm{temp}_s[\theta] - \mu_j[\theta])^2}}{\left |\mu_j[\theta]\right |\sqrt{j}};$
    }\\
    
    Break the loop when $ \delta_j[\theta]<tol$ for each $\theta$;
}
\Return{$\partial\mathbb{P}/\partial \theta=\mu_j[\theta]\; \mathrm{for\; each}\; \theta$, $N=j$}
\end{algorithm}

It is worth noting that, when a large number of design parameters are considered, an equal number of impulse response sensitivity time-history analyses need to be conducted in Step ii). This initial setup cost can become computationally expensive for large-scale linear systems. To address this issue, the adjoint variable method developed in \cite{hu2016explicit} can be employed to reduce the computational cost, in which the numerous impulse response sensitivity time-history analyses based on the equation of motion are replaced by a single impulse response time-history analysis governed by an adjoint equation that has the same form as the original motion equation. Therefore, owing to the merit of the adjoint variable method, the initial setup cost of the surface decomposition method is also not associated with the number of design parameters.

\section{Numerical examples}
In this section, three numerical examples, including an oscillator, a shear-type structure and a building frame structure, are investigated to demonstrate the effectiveness of the proposed surface decomposition method (SDM) for sensitivity analysis of first-passage dynamic reliability of linear systems. The first-passage dynamic reliability analysis of the linear system is also conducted, utilizing the efficient importance sampling (ISEE) \cite{au2001first}, the domain decomposition method (DDM) \cite{katafygiotis2006domain} and the directional importance sampling (DIS) \cite{misraji2020application}. The DIS is also employed to perform the reliability sensitivity analysis \cite{valdebenito2021sensitivity}. The finite difference method in conjunction with the importance sampling, which can be denoted as FDM-IS, is adopted to obtain the reference solutions to the sensitivities of failure probabilities. The basic idea of ISEE and FDM-IS are presented in \ref{appendix a} and \ref{appendix b}, respectively.

\subsection{An oscillator}\label{exam 1}
Consider the following equation of motion for a linear oscillator as 
\begin{equation}\label{oscillator equation}
     \ddot{u}(\vect X ,t)+2\omega_{n}\zeta_{n}  \dot{u}(\vect X ,t)+\omega^2_n u(\vect X ,t)
    = -\ddot{u}_g(\vect X ,t) 
\end{equation}
where $\omega_n=4\pi\mathrm{rad/s}$ and $\zeta_n=0.05$ are the circular frequency and damping ratio of the linear oscillator, respectively; $\ddot{ u}(\vect X ,t)$, $\dot{u}(\vect X ,t)$ and $u(\vect X ,t)$ are the acceleration, velocity and displacement of the linear oscillator, respectively; and $\ddot{u}_g(\vect X ,t)$ is the random seismic excitation assumed to be a stationary Gaussian white noise described by a power spectrum $S=5.5\times 10^{-4}\mathrm{m}^2/\mathrm{s}^3$. For spectral representation method, the upper and lower cutoff circular frequencies are $\omega_{\max}=25\pi$ and $\omega_{\min}=0$, respectively, while the number of circular frequency intervals is set as $q=500$. This leads to a 1000-dimensional standard Gaussian random vector $\vect X$.

The failure event $F$ is defined as that the peak absolute displacement of the oscillator exceeds a prescribed symmetric threshold $c$ within a given time duration $T=20\mathrm{s}$, which is expressed as
\begin{equation}\label{example1}
     F=\left\{c- \max_{t\in[0,T]} \left | u(\vect X ,t)\right | \leq 0\right\}
\end{equation}
The time step is set as $\Delta t=0.02\mathrm{s}$, resulting in a total of $n=T/\Delta t=1000$ component failure events.

\begin{figure}[ht]
  \centering
  \captionsetup{
  }
  \includegraphics[width=1\textwidth] {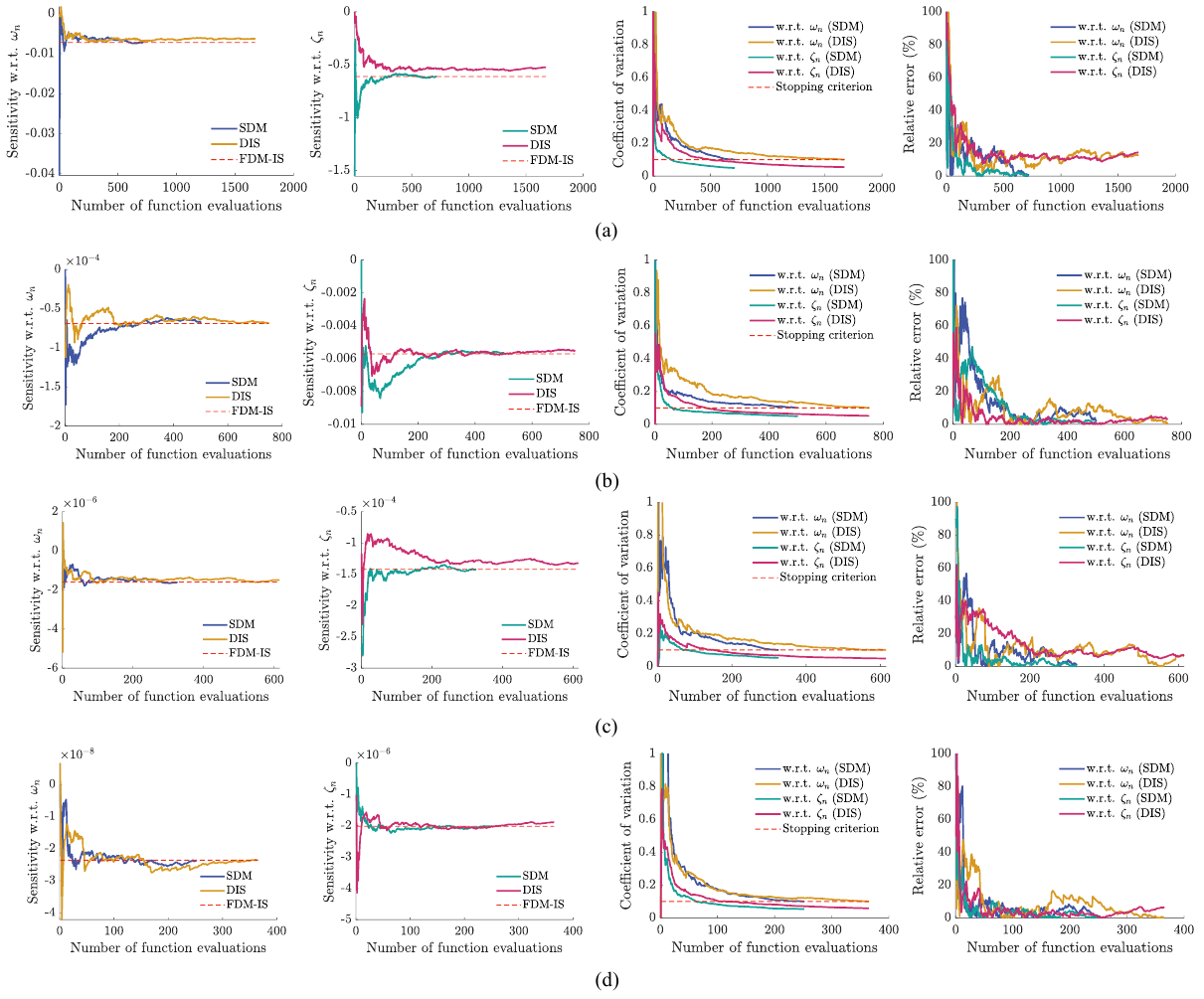}
  \caption{\textbf{Convergence histories of sensitivities of failure probability with respect to $\omega_n$ and $\zeta_n$ for Example 1.} (a) Case 1: $c=0.013\mathrm{m}$; (b) Case 2: $c=0.016\mathrm{m}$; (c) Case 3: $c=0.018\mathrm{m}$; (d) Case 4: $c=0.020\mathrm{m}$. The number of function evaluations required in the proposed SDM for Case 1, Case 2, Case 3 and Case 4 are 714, 501, 326 and 252, respectively, while those required in DIS are 1673, 751, 615, 365, respectively. Accordingly, the speedup ratios of the proposed SDM over DIS for these four cases are approximately 2.34, 1.50, 1.89 and 1.45, respectively.}  
  \label{fig:2}
\end{figure}

The circular frequency $\omega_n$ and damping ratio $\zeta_n$ of the oscillator are selected as the design parameters. To investigate the effect of the failure probability magnitude on the efficiency of the proposed method for sensitivity analysis of dynamic reliability, four cases are considered for the symmetric threshold, i.e., $c=0.013\mathrm{m}$, $c=0.016\mathrm{m}$, $c=0.018\mathrm{m}$, and $c=0.020\mathrm{m}$. The results of failure probabilities for these four cases, obtained from ISEE, DDM and DIS with a target coefficient of variation (COV) of 0.1, are presented in Table~\ref{tab:failure_probability Example 1}. The corresponding sensitivities of the failure probability with respect to $\omega_n$ and $\zeta_n$, computed by the proposed SDM and the DIS with a target COV of 0.1, are shown in Table~\ref{tab:failure_sensitivity for example 1}, together with the reference solutions obtained from FDM-IS using a 0.1\% perturbation in the design parameters and a target COV of 0.02. It is noted that, since both SDM and DIS are able to reuse the results of function evaluations when estimating the sensitivities of failure probability with respect to different design parameters, only a single number of function evaluations is reported for SDM and DIS in Table~\ref{tab:failure_sensitivity for example 1}. However, for FDM-IS, two separate numbers of function evaluations are given for the sensitivity estimation with respect to $\omega_n$ and $\zeta_n$, respectively. The convergence histories of the sensitivities of failure probability, evaluated by SDM and DIS for all cases, together with the corresponding COVs and the relative errors with respect to the reference solutions, are depicted in Figure \ref{fig:2}.

\begin{table}
\centering
\captionsetup{
  }
\footnotesize
\caption{Results of failure probabilities for Example~1.}
\label{tab:failure_probability Example 1}
\renewcommand{\arraystretch}{1.2}
\setlength{\tabcolsep}{4pt}
\begin{tabular}{c c c c c c}
\hline
Case & $c$ (m) & Method & $\mathbb{P}$ & Function Evaluations & COV \\
\hline
\multirow{3}{*}{1} & \multirow{3}{*}{0.013}
& ISEE & $3.06\times10^{-3}$ & 35 & 0.099 \\
& & DDM  & $3.32\times10^{-3}$ & 32 & 0.099 \\
& & DIS  & $3.28\times10^{-3}$ & 28 & 0.097 \\
\hline
\multirow{3}{*}{2} & \multirow{3}{*}{0.016}
& ISEE & $1.92\times10^{-5}$ & 30 & 0.097 \\
& & DDM  & $2.06\times10^{-5}$ & 29 & 0.099 \\
& & DIS  & $1.94\times10^{-5}$ & 21 & 0.099 \\
\hline
\multirow{3}{*}{3} & \multirow{3}{*}{0.018}
& ISEE & $3.85\times10^{-7}$ & 23 & 0.098 \\
& & DDM  & $3.67\times10^{-7}$ & 24 & 0.099 \\
& & DIS  & $3.63\times10^{-7}$ & 15 & 0.093 \\
\hline
\multirow{3}{*}{4} & \multirow{3}{*}{0.020}
& ISEE & $4.01\times10^{-9}$ & 20 & 0.096 \\
& & DDM  & $4.04\times10^{-9}$ & 18 & 0.099 \\
& & DIS  & $4.26\times10^{-9}$ & 11 & 0.081 \\
\hline
\end{tabular}
\end{table}

\begin{table}
\centering
\captionsetup{
  }
\footnotesize
\caption{Results of sensitivities of failure probability with respect to $\omega_n$ and $\zeta_n$ for Example~1.}
\label{tab:failure_sensitivity for example 1}
\renewcommand{\arraystretch}{1.2}
\setlength{\tabcolsep}{4pt}
\begin{tabular}{c c c c c c c c}
\hline
Case & $c$ (m) & Method & $\partial \mathbb{P}/\partial\omega_n$ & COV  & $\partial \mathbb{P}/\partial\zeta_n$ & COV & Function Evaluations \\
\hline
\multirow{3}{*}{1} & \multirow{3}{*}{0.013}
& SDM    & $-7.30\times10^{-3}$ & 0.100 & $-6.14\times10^{-1}$ & 0.049 & 714 \\
& & DIS    & $-6.32\times10^{-3}$ & 0.100 & $-5.24\times10^{-1}$ & 0.054 & 1673 \\
& & FDM-IS & $-7.21\times10^{-3}$ & 0.020 & $-6.11\times10^{-1}$ & 0.020 & $6.17\times10^{5}/1.28\times10^{6}$ \\
\hline
\multirow{3}{*}{2} & \multirow{3}{*}{0.016}
& SDM    & $-6.80\times10^{-5}$ & 0.100 & $-5.67\times10^{-3}$ & 0.049 & 501 \\
& & DIS    & $-6.96\times10^{-5}$ & 0.100 & $-5.70\times10^{-3}$ & 0.051 & 751 \\
& & FDM-IS & $-6.90\times10^{-5}$ & 0.020 & $-5.71\times10^{-3}$ & 0.020 & $2.96\times10^{5}/6.60\times10^{5}$ \\
\hline
\multirow{3}{*}{3} & \multirow{3}{*}{0.018}
& SDM    & $-1.62\times10^{-6}$ & 0.100 & $-1.42\times10^{-4}$ & 0.052 & 326 \\
& & DIS    & $-1.58\times10^{-6}$ & 0.100 & $-1.43\times10^{-4}$ & 0.048 & 615 \\
& & FDM-IS & $-1.60\times10^{-6}$ & 0.020 & $-1.42\times10^{-4}$ & 0.020 & $1.98\times10^{5}/4.65\times10^{5}$ \\
\hline
\multirow{3}{*}{4} & \multirow{3}{*}{0.020}
& SDM    & $-2.40\times10^{-8}$ & 0.100 & $-2.02\times10^{-6}$ & 0.053 & 252 \\
& & DIS    & $-2.25\times10^{-8}$ & 0.100 & $-1.92\times10^{-6}$ & 0.058 & 365 \\
& & FDM-IS & $-2.36\times10^{-8}$ & 0.020 & $-2.02\times10^{-6}$ & 0.020 & $1.36\times10^{5}/3.47\times10^{5}$ \\
\hline
\end{tabular}
\end{table}

It can be observed from Table~\ref{tab:failure_probability Example 1} that all three methods, namely ISEE, DDM and DIS, exhibit very high efficiency in computing the failure probabilities, with their efficiency increasing as the failure probability decreases. This phenomenon is consistent with the findings presented in \cite{au2001first,katafygiotis2006domain}. Interestingly, the proposed SDM and the DIS also possess this counterintuitive characteristic in the context of reliability sensitivity analysis, demonstrating higher efficiency as the failure probability becomes smaller, as shown in Table~\ref{tab:failure_sensitivity for example 1}. The mechanism underlying the above phenomena is that the efficiency of all these methods, whether applied to first-passage dynamic reliability analysis or reliability sensitivity analysis, is strongly dependent on the degree of overlap among the component failure domains/surfaces. A lower overlapping degree among the component failure domains/surfaces corresponds to higher efficiency of these methods. In general, a first-passage failure event with a lower probability implies larger distances between the origin and the component failure domains/surfaces, which results in a lower degree of overlap and finally leads to higher efficiency of these methods.

Overall, the sensitivity results obtained from SDM and DIS are in good agreement, and they also agree well with the reference solutions obtained from FDM-IS. Moreover, the SDM exhibits higher efficiency than DIS, requiring less than 1000 function evaluations for all cases, thereby demonstrating the effectiveness of the proposed method. Case 1 is taken for further illustration. The proposed SDM requires only 714 function evaluations to compute the sensitivities of failure probability with respect to $\omega_n$ and $\zeta_n$, while 1673 function evaluations are required in DIS meet the same stopping criterion. The speedup ratio of SDM over DIS is approximately 2.34. For the reference method FDM-IS, $1.36\times10^5$ and $3.47\times10^5$ function evaluations are needed to obtain the sensitivity results with respect to $\omega_n$ and $\zeta_n$, respectively.

\begin{figure}[ht]
  \centering
  \captionsetup{
  }
  \includegraphics[width=1\textwidth] {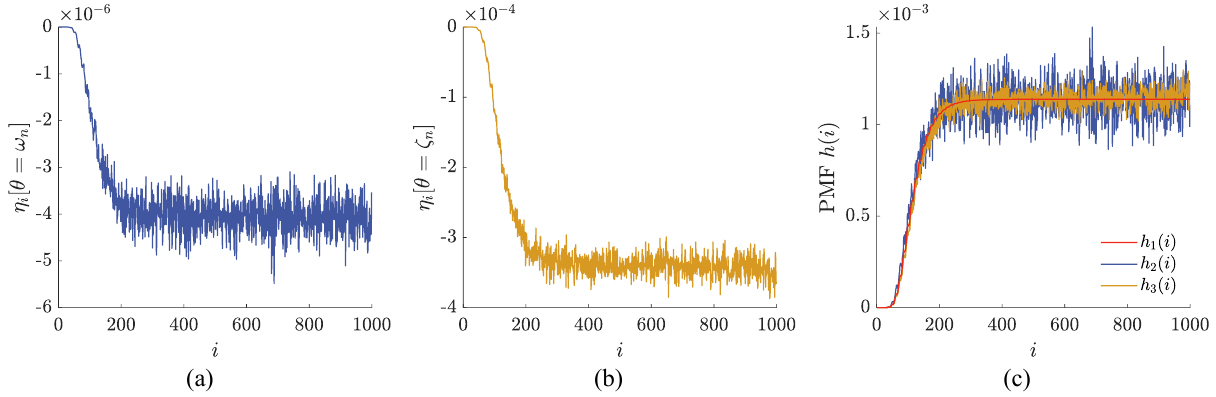}
  \caption{\textbf{Comparison among different PMFs.} (a) Estimation of component surface integrals with respect to $\omega_n$; (b) Estimation of component surface integrals with respect to $\zeta_n$; (c) Comparison among different PMFs.}  
  \label{fig:3}
\end{figure}

In the present SDM, the importance sampling PMF is chosen as $h_1(i)=\mathbb{P}_{i} / \sum_{s=1}^{n}\mathbb{P}_{s}$, where $\mathbb{P}_{i}\;(i=1,2,\dots,n)$ are the component failure probabilities. To demonstrate that this PMF provides a sufficient surrogate for the theoretically optimal PMF, the optimal PMFs with respect to $\omega_n$ and $\zeta_n$, denoted by $h_2(i)=\eta_{i}[\theta=\omega_n] / \sum_{s=1}^{n}\eta_{s}[\theta=\omega_n]$ and $h_3(i)=\eta_{i}[\theta=\zeta_n] / \sum_{s=1}^{n}\eta_{s}[\theta=\zeta_n]$, respectively, are also evaluated. Here, $\eta_{i}[\theta=\omega_n]$ and $\eta_{i}[\theta=\zeta_n]\;(i=1,2,\dots,n)$ denote the component surface integrals with respect to $\omega_n$ and $\zeta_n$, respectively, which are estimated by the direct Monte Carlo simulation, and the corresponding results are presented in Figures \ref{fig:3}(a) and \ref{fig:3}(b). The resulting optimal PMFs $h_2(i)$ and $h_3(i)$ as well as the adopted PMF $h_1(i)$ are depicted in Figure \ref{fig:3}(c), from which it can be seen that $h_1(i)$ serves as a good surrogate for the optimal PMFs. 
Notably, the fact that sensitivity estimations with respect to different design parameters share the same surrogate PMF is the key reason that enables the reuse of function evaluation results in SDM.

\begin{figure}[ht]
  \centering
  \captionsetup{
  }
  \includegraphics[width=1\textwidth] {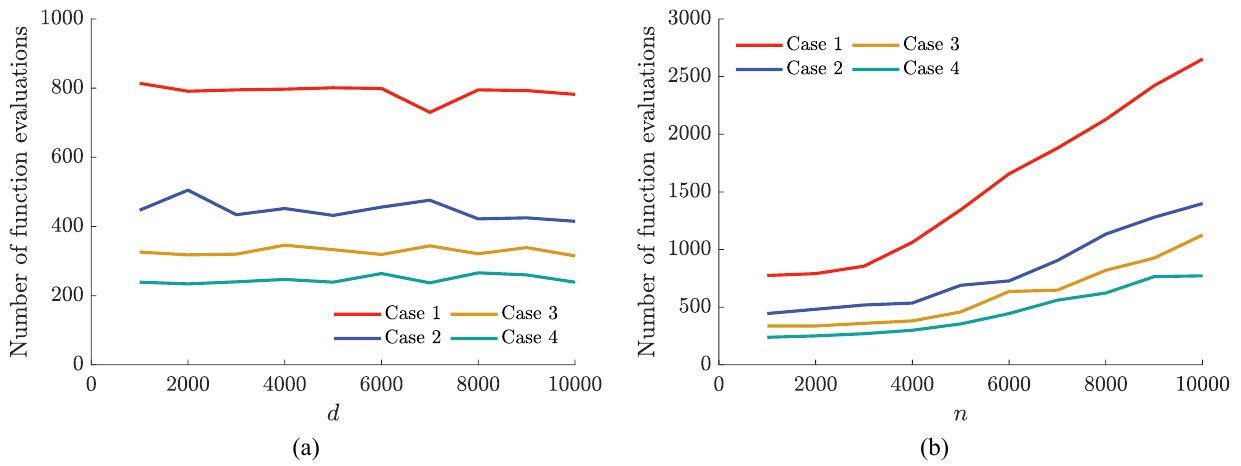}
  \caption{\textbf{Comparison of computational efficiency of SDM for different input dimensionalities $d$ and numbers of component failure events $n$.} (a) Number of function evaluations versus $d$, with $d\in[1000,10000]$ and fixed $n=1000$; (b) Number of function evaluations versus $n$, with $n\in[1000,10000]$ and fixed $d=1000$.}  
  \label{fig:4}
\end{figure}

To investigate the computational efficiency of SDM for higher input dimensionality and larger number of component failure events, the SDM is further employed to perform the reliability sensitivity analysis for different input dimensionalities $d$ and different numbers of component failure events $n$. Two scenarios are considered: (a) $d$ varies from 1000 to 10000 with $n=1000$ fixed; (b) $n$ varies from 1000 to 10000 with $d=1000$ fixed. Different values of $d$ and $n$ can be achieved by scaling the circular frequency interval $\Delta\omega$ and time duration $T$, respectively. The number of function evaluations required in SDM corresponding to the same target COV of 0.1 for these two scenarios are shown in Figures \ref{fig:4}(a) and \ref{fig:4}(b), respectively. It can be seen from Figure \ref{fig:4} that the efficiency of SDM is independent of the input dimensionality, while it decreases as the number of component failure events increases. This is expected since a larger number of samples are required to adequately account for the contributions from the additional component surface integrals to the total sensitivity estimation. It should be noted that, in practical engineering scenarios with nonstationary excitations, an increase in the number of component failure events does not necessarily imply a reduction in SDM efficiency, since only the component surface integrals corresponding to the strong-motion intervals contribute significantly to the total sensitivity estimation.

\subsection{A shear-type structure}
\begin{figure}
  \centering  \includegraphics[width=0.4\textwidth] {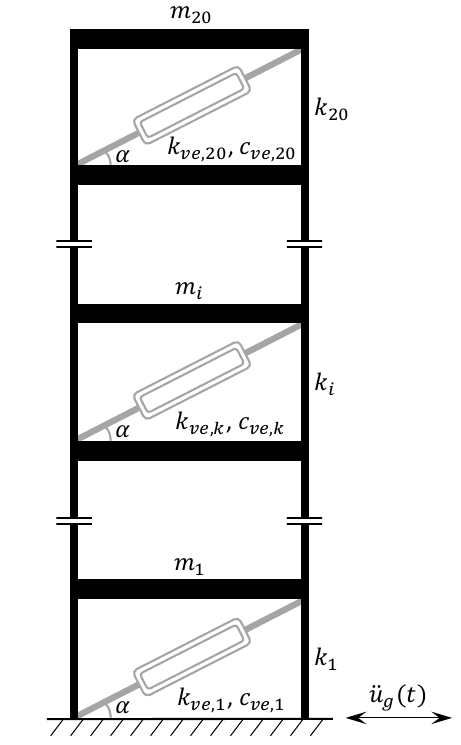}
  \caption{\color{Black}\textbf{A 20-degree-of-freedom shear-type structure equipped with viscoelastic dampers.} }  
  \label{fig:5}
\end{figure}

Consider a 20-degree-of-freedom shear-type structure, where each story is equipped with a viscoelastic damper, as illustrated in Figure \ref{fig:5}. The mass and stiffness of each story of the structure are $m_i=3\times10^3\mathrm{kg}$ and $k_i=3\times10^4\mathrm{kN/m}\;(i=1,2,\dots,20)$, respectively. The Rayleigh damping model is adopted to define the damping matrix, assuming a damping ratio of 0.05 for the 1st and 20th modes of the structure. The angle between the braced arm of the viscoelastic damper and the floor is $\alpha=\arccos0.8$. The mechanical behavior of the viscoelastic damper is characterized by the Kelvin model as follows:
\begin{equation}\label{Kelvin}
     f_{ve,k}(t) = k_{ve,k}u_{ve,k}(t)+ c_{ve,k}\dot{u}_{ve,k}(t)\quad(k=1,2,\dots,20)
\end{equation}
where $f_{ve,k}(t)$ is the damping force of the $k$th viscoelastic damper; $k_{ve,k}=3\times10^3\mathrm{kN/m}$ and $c_{ve,k}=2.5\times10^3\mathrm{kN\cdot s/m}\;(k=1,2,\dots,20)$ are the stiffness and damping coefficients of the $k$th viscoelastic damper, respectively; and $u_{ve,k}(t)$ and $\dot{u}_{ve,k}(t)$ are the axial nodal relative displacement and velocity between the two ends of the $k$th viscoelastic damper, respectively.

The structure is subjected to a ground motion acceleration $\ddot{u}_g(\vect X,t)$, which is assumed to be a uniformly modulated nonstationary Gaussian random process, i.e., $\ddot{u}_g(\vect X,t)=g(t)\ddot{u}^0_g(\vect X,t)$. $g(t)$ is the modulation function taken as
\begin{equation}\label{modulation function}
g(t)=
\begin{cases}
\left( t/t_a \right)^2, & 0 \le t \le t_a, \\[6pt]
1, & t_a \le t \le t_b, \\[6pt]
e^{-\lambda(t-t_b)}, & t_b \le t \le t_c.
\end{cases}
\end{equation}
where $t_a=8\mathrm{s}$, $t_b=20\mathrm{s}$, $t_c=30\mathrm{s}$ and $\lambda=0.1572$. $\ddot{u}^0_g(\vect X,t)$ is the corresponding stationary Gaussian random process, with its correlation function expressed as
\begin{equation}\label{correlation function 1}
R_{\ddot{u}_g^{0}}(\tau) = \frac{\pi S_0}{2} 
e^{-\zeta_g \omega_g |\tau|}
\left( \mu_1 \cos \omega_d \tau + \mu_2 \sin \omega_d |\tau| \right)
\end{equation}
where $\tau$ denotes the time lag; $S_0$ is the intensity parameter directly associated with the mean peak ground acceleration; $\omega_g=14\mathrm{rad/s}$ and $\zeta_g=0.6$; and $\omega_d $, $\mu_1$ and $\mu_2$ are given by
\begin{equation}\label{correlation function 2}
\omega_d = \omega_g \sqrt{1 - \zeta_g^2}, \quad
\mu_1 = \frac{\omega_g (1 + 4\zeta_g^2)}{\zeta_g}, \quad
\mu_2 = \frac{\omega_g (1 - 4\zeta_g^2)}{\sqrt{1 - \zeta_g^2}}.
\end{equation}
Consequently, the correlation function of $\ddot{u}_g(\vect X,t)$ can be expressed as
\begin{equation}\label{correlation function 3}
R_{\ddot{u}_g}(t,\tau)=g(t)g(t+\tau)R_{\ddot{u}_g^{0}}(\tau)
\end{equation}

For orthogonal decomposition method, the time step and time duration of the ground motion acceleration are taken as $\Delta t=0.02\mathrm{s}$ and $T=30\mathrm{s}$, respectively, and hence the number of time steps is $n=T/\Delta t=1500$. This results in a 1500-dimensional standard Gaussian random vector $\vect X$.

The failure event $F$ is defined as that the maximum absolute interstory drift among the 20 stories exceeds a given symmetric threshold $c=0.006\mathrm{m}$ within the time duration $T=30\mathrm{s}$, which can be expressed as
\begin{equation}\label{example2}
     F=\left\{c- \max_{i \in \left\{ 1,2,\dots,20 \right\}} \max_{t\in[0,T]} \left | u_i(\vect X ,t)\right | \leq 0\right\}
\end{equation}
where $u_i(\vect X ,t)\;(i=1,2,\dots,20)$ is the interstory drift of the $i$th story. Since the number of time steps is $n=1500$ and the number of critical response components is $m=20$, a total of $mn=30000$ component failure events are considered in this example.

\begin{figure}
  \centering
  \includegraphics[width=0.8\textwidth] {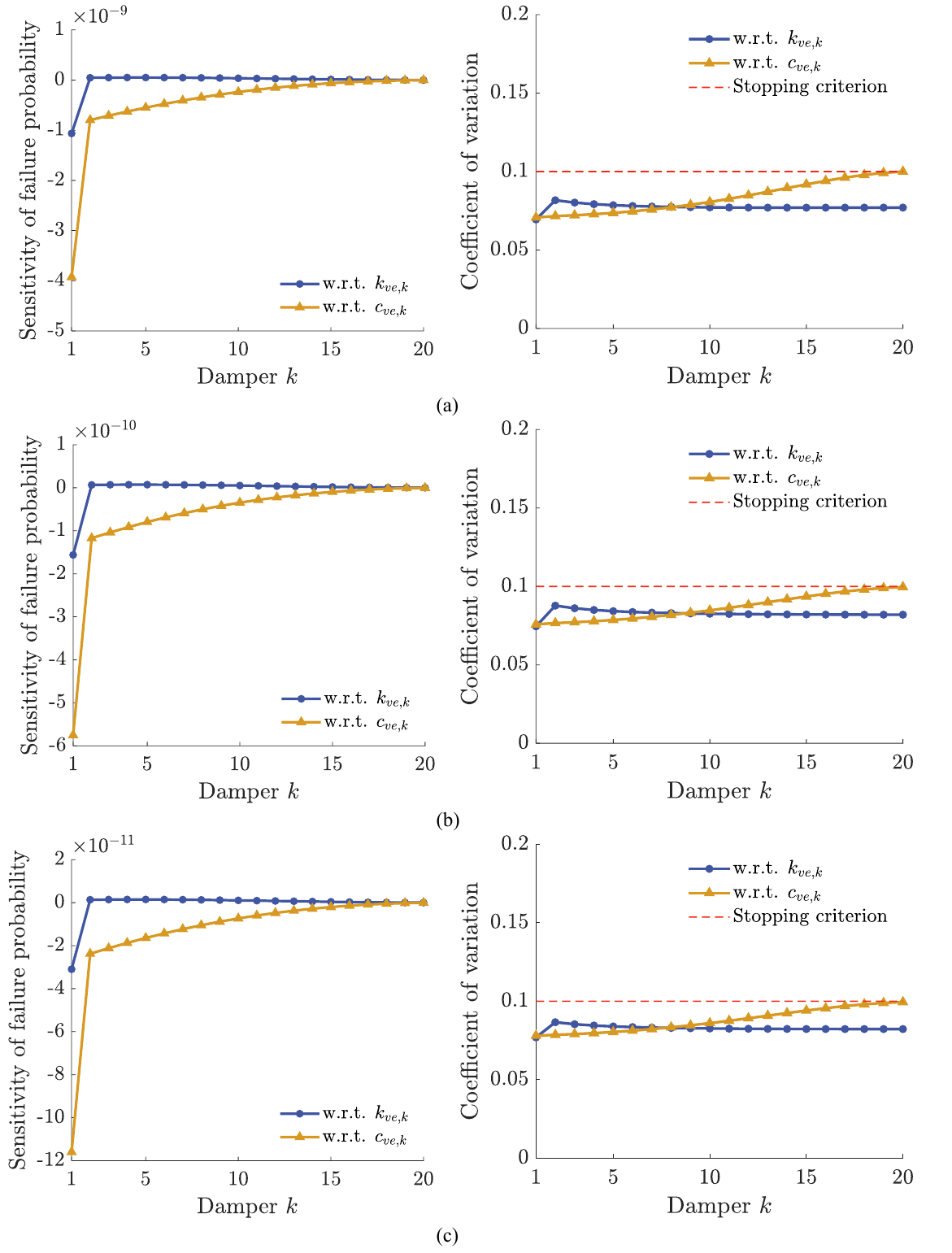}
  \caption{\color{Black}\textbf{Results of sensitivities of failure probability with respect to $k_{ve,k}$ and $c_{ve,k}\;(k=1,2,\dots,20)$ for Example 2.} (a) Case 1: $S_0=0.010\mathrm{m^2/s^3}$; (b) Case 2: $S_0=0.008\mathrm{m^2/s^3}$; (c) Case 3: $S_0=0.007\mathrm{m^2/s^3}$. The number of function evaluations required in the proposed SDM for Case 1, Case 2 and Case 3 are 1663, 953 and 877, respectively.}  
  \label{fig:6}
\end{figure}

The stiffness and damping coefficients of the viscoelastic dampers, $k_{ve,k}$ and $c_{ve,k}\;(k=1,2,\dots,20)$, are selected as the design parameters. Three cases are considered for the intensity parameter, i.e., $S_0=0.010\mathrm{m^2/s^3}$, $S_0=0.008\mathrm{m^2/s^3}$ and $S_0=0.007\mathrm{m^2/s^3}$, which leads to different magnitudes of the failure probability. The results of failure probabilities are presented in Table \ref{tab:failure_probability Example 2}, from which it can be observed that all three methods, including ISEE, DDM and DIS, exhibit very high efficiency for the first-passage dynamic reliability analysis, and DIS is the most efficient among them. The results of sensitivities of failure probability with respect to all design parameters $k_{ve,k}$ and $c_{ve,k}\;(k=1,2,\dots,20)$ are depicted in Figure \ref{fig:6}, which are obtained from the proposed SDM with a target COV of 0.1 using 1663, 953 and 877 function evaluations for Case 1, Case 2 and Case 3, respectively. Consistent with the observation in Example 1, SDM exhibits higher computational efficiency as the failure probability decreases. It is seen from Figure \ref{fig:6} that the failure probability of the structure is most sensitive to the parameters of the first-story viscoelastic damper, i.e., $k_{ve,1}$ and $c_{ve,1}$, while the sensitivity computation with respect to the damping coefficient of the top-story viscoelastic damper, i.e., $c_{ve,20}$, dominates the convergence behavior of SDM.

\begin{table}[htbp]
\centering
\captionsetup{
}
\footnotesize
\caption{Results of failure probabilities for Example~2.}
\label{tab:failure_probability Example 2}
\renewcommand{\arraystretch}{1.2}
\setlength{\tabcolsep}{4pt}
\begin{tabular}{c c c c c c}
\hline
Case & $S_0$ (m$^2$/s$^3$) & Method & $\mathbb{P}$ & Function Evaluations & COV \\
\hline
\multirow{3}{*}{1} & \multirow{3}{*}{0.010}
& ISEE & $3.83\times10^{-3}$ & 188 & 0.100 \\
& & DDM  & $3.79\times10^{-3}$ & 183 & 0.100 \\
& & DIS  & $3.49\times10^{-3}$ & 100 & 0.100 \\
\hline
\multirow{3}{*}{2} & \multirow{3}{*}{0.008}
& ISEE & $3.63\times10^{-4}$ & 142 & 0.099 \\
& & DDM  & $3.56\times10^{-4}$ & 139 & 0.099 \\
& & DIS  & $3.82\times10^{-4}$ & 79 & 0.098 \\
\hline
\multirow{3}{*}{3} & \multirow{3}{*}{0.007}
& ISEE & $7.04\times10^{-5}$ & 95 & 0.099 \\
& & DDM  & $6.77\times10^{-5}$ & 84 & 0.099 \\
& & DIS  & $6.72\times10^{-5}$ & 66 & 0.099 \\
\hline
\end{tabular}
\end{table}

\begin{figure}
  \centering
  \captionsetup{
  }
  \includegraphics[width=1\textwidth] {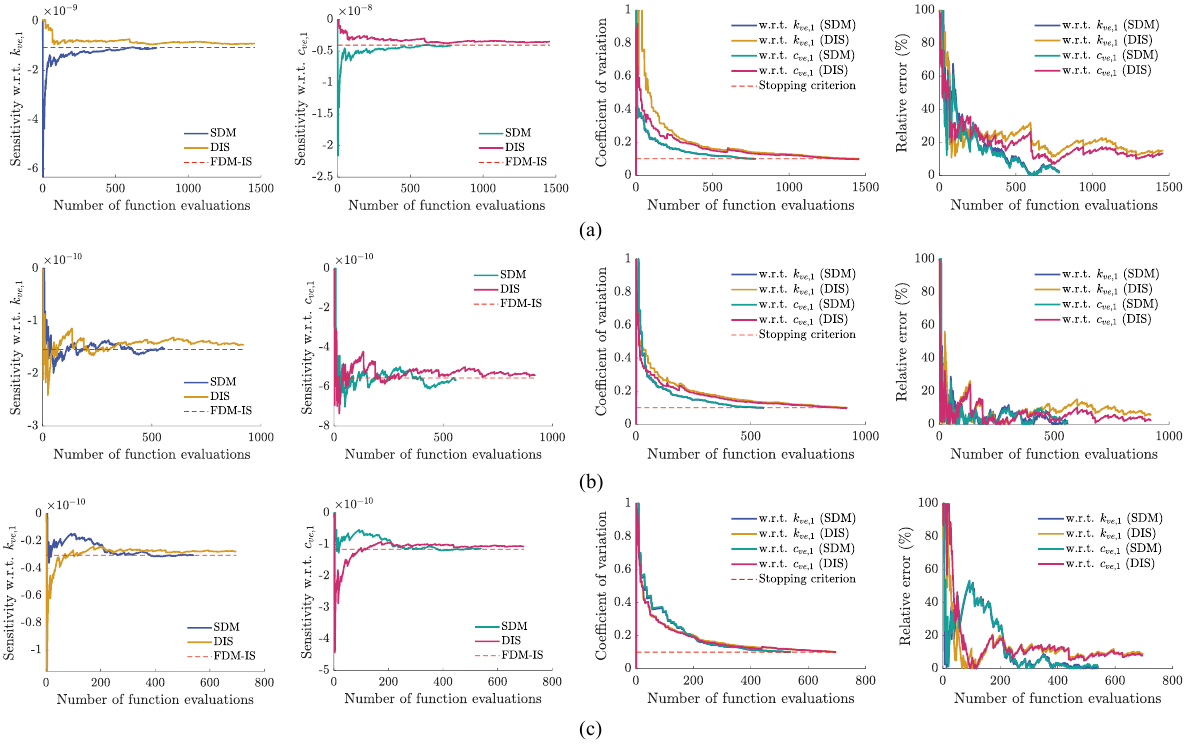}
  \caption{\textbf{Convergence histories of sensitivities of failure probability with respect to $k_{ve,1}$ and $c_{ve,1}$ for Example 2.} (a) Case 1: $S_0=0.010\mathrm{m^2/s^3}$; (b) Case 2: $S_0=0.008\mathrm{m^2/s^3}$; (c) Case 3: $S_0=0.007\mathrm{m^2/s^3}$. The number of function evaluations required in the proposed SDM for Case 1, Case 2 and Case 3 are 782, 558 and 540, respectively, whereas those required in DIS are 1458, 920 and 697, respectively. The speedup ratios of SDM over DIS for these three cases are approximately 1.86, 1.65 and 1.29, respectively.}  
  \label{fig:7}
\end{figure}

The sensitivities of the failure probability with respect to $k_{ve,1}$ and $c_{ve,1}$ are further investigated using SDM and DIS with a target COV of 0.1, as summarized in Table \ref{tab:failure_sensitivity for example 2}. The reference solutions obtained from FDM-IS, with a target COV of 0.1 and a 0.1\% perturbation of design parameter, are also provided for comparison. The convergence histories of these sensitivities as well as the corresponding COVs and the errors relative to the reference solutions are illustrated in Figure \ref{fig:7}. It can be observed that the sensitivity results obtained from the proposed SDM and the DIS are in good agreement and both align well with the reference solutions. As for the efficiency, SDM outperforms DIS with speedup ratios of about 1.86, 1.65 and 1.29 for Case 1, Case 2 and Case 3, respectively. Overall, the proposed SDM demonstrates excellent accuracy and efficiency for the sensitivity analysis of first-passage dynamic reliability of linear systems.

\begin{table}
\centering
\captionsetup{
}
\footnotesize
\caption{Results of sensitivities of failure probability with respect to $k_{ve,1}$ and $c_{ve,1}$ for Example~2.}
\label{tab:failure_sensitivity for example 2}
\renewcommand{\arraystretch}{1.2}
\setlength{\tabcolsep}{4pt}
\begin{tabular}{c c c c c c c c}
\hline
Case & $S_0\;(\mathrm{m^2/s^3})$ & Method & $\partial \mathbb{P}/\partial k_{ve,1}$ & COV  & $\partial \mathbb{P}/\partial c_{ve,1}$ & COV & Function Evaluations \\
\hline
\multirow{3}{*}{1} & \multirow{3}{*}{0.010} 
& SDM & $-1.10\times10^{-9}$ & 0.099 & $-4.17\times10^{-9}$ & 0.100 & 782 \\
& & DIS & $-0.92\times10^{-9}$ & 0.100 & $-3.53\times10^{-9}$ & 0.097 & 1458 \\
& & FDM-IS & $-1.07\times10^{-9}$ & 0.100 & $-4.05\times10^{-9}$ & 0.100 & $1.84\times10^{6}/5.78\times10^{5}$ \\
\hline
\multirow{3}{*}{2} & \multirow{3}{*}{0.008} 
& SDM & $-1.54\times10^{-10}$ & 0.098 & $-5.70\times10^{-10}$ & 0.100 & 558 \\
& & DIS & $-1.46\times10^{-10}$ & 0.100 & $-5.46\times10^{-10}$ & 0.097 & 920 \\
& & FDM-IS & $-1.54\times10^{-10}$ & 0.100 & $-5.56\times10^{-10}$ & 0.100 & $1.05\times10^{6}/3.52\times10^{5}$ \\
\hline
\multirow{3}{*}{3} & \multirow{3}{*}{0.007} 
& SDM & $-3.08\times10^{-11}$ & 0.098 & $-1.15\times10^{-10}$ & 0.100 & 540 \\
& & DIS & $-2.77\times10^{-11}$ & 0.100 & $-1.06\times10^{-10}$ & 0.100 & 697 \\
& & FDM-IS & $-3.04\times10^{-11}$ & 0.100 & $-1.16\times10^{-10}$ & 0.100 & $9.26\times10^{5}/2.92\times10^{5}$ \\
\hline
\end{tabular}
\end{table}

\subsection{A building frame structure}
\begin{figure}
  \centering
  \captionsetup{
  }
  \includegraphics[width=1\textwidth] {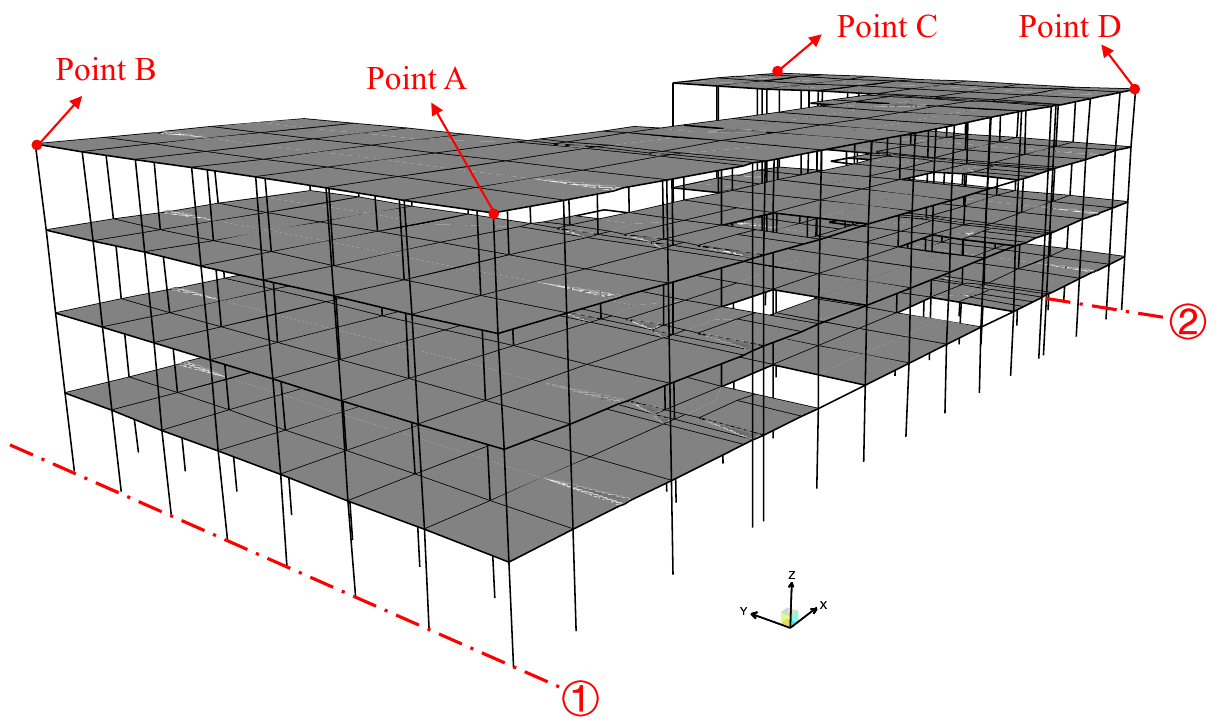}
  \caption{\textbf{A 4-storey reinforced concrete  building frame structure.}}  
  \label{fig:8}
\end{figure}

\begin{figure}
  \centering
  \captionsetup{
  }
  \includegraphics[width=1\textwidth] {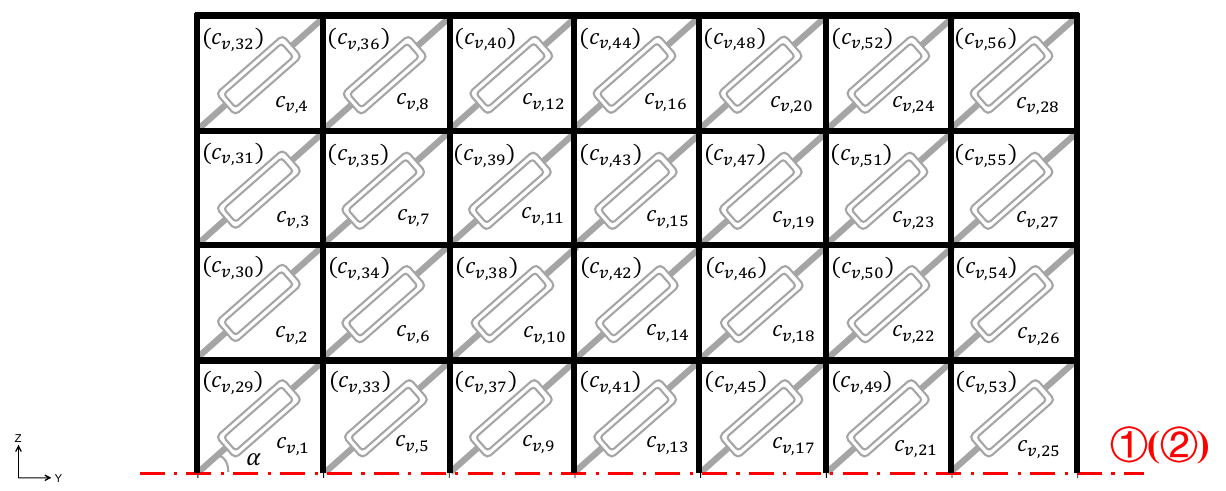}
  \caption{\textbf{Layouts of linear viscous dampers for planar frames  {\large\ding{172}} and  {\large\ding{173}}.}}  
  \label{fig:9}
\end{figure}

Consider a 4-storey reinforced concrete building frame structure with a storey height of 4.5m, as shown in Figure \ref{fig:8}. The two planar frames {\large\ding{172}} and  {\large\ding{173}} located on both sides of the building each consist of 7 bays with a bay width of 4.9m. The columns and beams for the frame structure are designed with rectangular cross sections measuring $0.4\mathrm{m}\times0.5\mathrm{m}$ and $0.8\mathrm{m}\times0.25\mathrm{m}$, respectively. The damping matrix is defined using the Rayleigh damping model, with a critical damping ratio of 0.05 assigned to the first and fifteenth modes of the structure. To
suppress the excessive displacements of the structure under the Y-direction
seismic excitation, each of the planar frames {\large\ding{172}} and  {\large\ding{173}} is equipped with 28 linear viscous dampers, as illustrated in Figure \ref{fig:9}.  The dampers in frames {\large\ding{172}} and {\large\ding{173}} are characterized by the damping coefficients $c_{v,1},c_{v,2},\dots,c_{v,28}$ and $c_{v,29},c_{v,30},\dots,c_{v,56}$, respectively. The angle between the braced
arm of the viscous damper and the floor is $\alpha=\arctan(4.5/4.9)$. The finite element model of the structure is developed in SAP2000, in which 1201 frame elements are used to model the columns and beams, and 530 shell elements are employed to simulate the floor slabs, resulting in a total of 4020 degrees of freedom.

The structure is subjected to a uniformly modulated non-stationary Gaussian ground motion acceleration $\ddot{u}_g(\vect X,t)=g(t)\ddot{u}_g^0(\vect X,t)$, in which the intensity modulation function $g(t)$ and the correlation function of the stationary part $\ddot{u}_g^0(\vect X,t)$ are the same as those defined in Eqs.\eqref{modulation function} and \eqref{correlation function 1}, respectively, except that the intensity parameter is specified as $S_0=3.5\times10^{-3}\mathrm{m}^2/\mathrm{s}^3$ in this example. The orthogonal decomposition method is also adopted to represent the seismic excitation, in which the time step and time duration are taken as $\Delta t=0.02\mathrm{s}$ and $T=30\mathrm{s}$, respectively, leading to $n=1500$ time steps. Therefore, the dimension of the standard Gaussian random vector $\vect X$ is 1500.

The failure event $F$ is defined as the exceedance of a prescribed symmetric threshold $c=0.01\mathrm{m}$ by 
the maximum absolute displacement among the 4 corner points A, B, C and D (see Figure \ref{fig:9}) over a time duration of $T=30\mathrm{s}$. The failure event $F$ can be expressed as
\begin{equation}\label{example3}
F = \left\{\, c - \max_{i \in \{\mathrm{A,B,C,D}\}} \max_{t\in[0,T]} |u_i(\vect X,t)| \le 0 \,\right\}
\end{equation}
where $u_i(\vect X,t)\;(i\in \{\mathrm{A,B,C,D}\})$ are the Y-direction horizontal displacements at points A, B, C and D, respectively. In this example, since the number of time steps is $n=1500$ and the number of critical response components is $m=4$, a total of $mn=6000$ component failure events are considered.

The damping coefficients of all the linear viscous dampers are selected as the design parameters, and they are set as $c_{v,k}=200\mathrm{kN\cdot s/m}\;(k=1,2,\dots,56)$. At the initial setup stage of SDM, the explicit expressions of the critical responses as well as their sensitivities with respect to all design parameters $c_{v,k}\;(k=1,2,\dots,56)$ need to be constructed. In principle, this would require a response time-history analysis plus 56 response sensitivity time-history analyses of the building frame structure. To reduce the computational cost, the adjoint variable method developed in \cite{hu2016explicit} is employed, where the 56 response sensitivity time-history analyses are replaced by solving the adjoint equations corresponding to the 4 critical responses. This is equivalent to performing 4 response time-history analyses of the structure. It should be noted that the initial setup cost of SDM, which now amounts to 5 response time-history analyses of the structure, remains almost unchanged even when additional design parameters are introduced.

\begin{figure}
  \centering
  \captionsetup{
  }
  \includegraphics[width=1\textwidth] {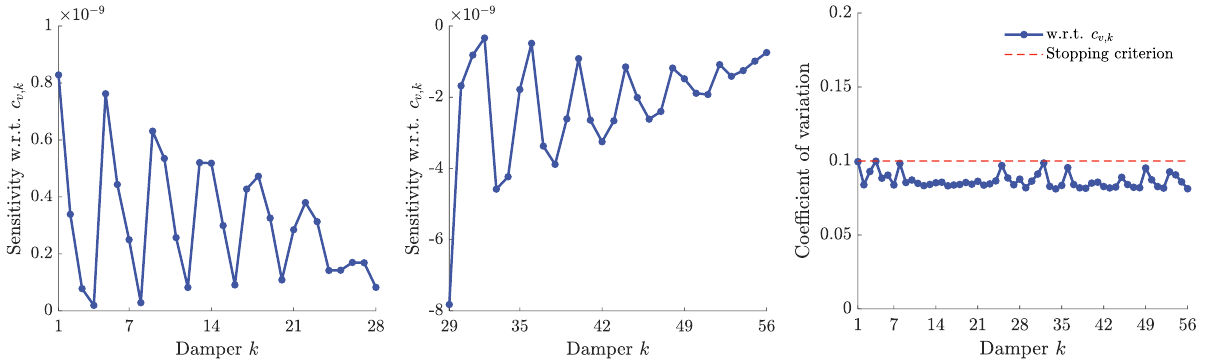}
  \caption{\textbf{Results of sensitivities of failure probability with respect to $c_{v,k}\;(k=1,2,\dots,56)$ for Example 3.} The number of function evaluations required in the proposed SDM is 502.}  
  \label{fig:10}
\end{figure}

The results of failure probabilities obtained from ISEE, DDM and DIS are presented in Table \ref{tab:failure_probability_example3}, from which it can be observed that all three methods demonstrate very high efficiency for computing the first-passage failure probability, with DIS achieving the best performance. The results of the failure probability sensitivities with respect to all design parameters $c_{v,k}\;(k=1,2,\dots,56)$, which are obtained through the proposed SDM using 502 function evaluations with a target COV of 0.1, are illustrated in Figure \ref{fig:10}. It can be observed that the failure probability sensitivities with respect to $c_{v,k}\;(k=1,2,\dots,28)$ are positive, while those with respect to $c_{v,k}\;(k=29,30,\dots,56)$ are negative, meaning that increasing the damping coefficients of the dampers installed on frame {\large\ding{172}} leads to an increase in the failure probability, whereas increasing the damping coefficients of the dampers on frame {\large\ding{173}} reduces the failure probability. This is reasonable when the system failure is governed by the displacement responses at points C and D on frame {\large\ding{173}}. In this case, increasing the damping on frame {\large\ding{172}} and frame {\large\ding{173}} may result in completely different effects on the displacement responses of frame {\large\ding{173}}, thereby leading to opposite trends in the failure probability. 

It can be also observed from Figure \ref{fig:10} that the failure probability is more sensitive to the damping coefficients of the dampers on the first storey than to those on the upper storeys. Furthermore, the failure probability exhibits the highest positive sensitivity to the damping coefficient $c_{v,1}$ and the highest negative sensitivity to the damping coefficient $c_{v,29}$. With the results of failure probability and the corresponding sensitivities at hand, certain gradient-based optimizers, such as the method of moving asymptotes \cite{svanberg1987method}, can be employed to update the design parameters $c_{v,k}\;(k=1,2,\dots,56)$. This process is typically repeated within the framework of reliability-based design optimization until convergence of the design parameters is achieved. 

\begin{table}
\centering
\captionsetup{
}
\footnotesize
\caption{Results of failure probabilities for Example 3.}
\label{tab:failure_probability_example3}
\renewcommand{\arraystretch}{1.2}
\setlength{\tabcolsep}{6pt}
\begin{tabular}{c c c c}
\hline
Method & $\mathbb{P}$ & Function Evaluations & COV \\
\hline
ISEE & $1.76\times10^{-3}$ & 94 & 0.099 \\
DDM  & $1.73\times10^{-3}$ & 84 & 0.099 \\
DIS  & $1.74\times10^{-3}$ & 59 & 0.099 \\
\hline
\end{tabular}
\end{table}

\begin{figure}
  \centering
  \captionsetup{
  }
  \includegraphics[width=1\textwidth] {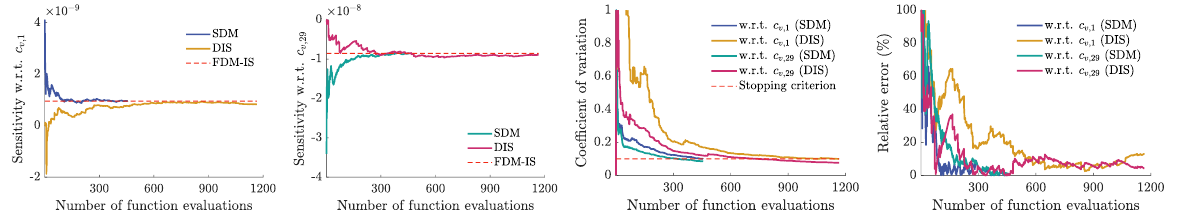}
  \caption{\textbf{Convergence histories of sensitivities of failure probability with respect to $c_{v,1}$ and $c_{v,29}$ for Example 3.} The number of function evaluations required in the proposed SDM and the DIS are 456 and 1166, respectively. The speedup ratio of SDM over DIS is approximately 2.56.}  
  \label{fig:11}
\end{figure}

The sensitivities of the failure probability with respect to $c_{v,1}$ and $c_{v,29}$ are further explored using SDM, DIS and FDM-IS (served as reference solutions) with a target COV of 0.1. The results are presented in Table \ref{tab:failure_sensitivity_example3}, while the convergence histories of the failure probability sensitivities, along with the corresponding COVs and errors relative to the reference solutions, are depicted in Figure \ref{fig:11}. It can be observed that both SDM and DIS exhibit high accuracy, while SDM outperforms DIS in terms of efficiency with a speedup ratio of 2.56. Overall, the proposed SDM demonstrates its potential for solving practical engineering problems.

\begin{table}
\centering
\captionsetup{
}
\footnotesize
\caption{Results of sensitivities of failure probability with respect to $c_{v,1}$ and $c_{v,29}$ for Example 3.}
\label{tab:failure_sensitivity_example3}
\renewcommand{\arraystretch}{1.2}
\setlength{\tabcolsep}{6pt}
\begin{tabular}{c c c c c c}
\hline
Method 
& $\partial \mathbb{P}/\partial c_{v,1}$ & COV  
& $\partial \mathbb{P}/\partial c_{v,29}$ & COV 
& Function Evaluations \\
\hline
SDM     & $9.68\times10^{-10}$ & 0.100 & $-8.68\times10^{-9}$ & 0.085 & 456 \\
DIS     & $8.30\times10^{-10}$ & 0.100 & $-8.90\times10^{-9}$ & 0.083 & 1166 \\
FDM-IS  & $9.54\times10^{-10}$ & 0.100 & $-8.55\times10^{-9}$ & 0.100 & $8.93\times10^{6}/2.12\times10^{6}$ \\
\hline
\end{tabular}
\end{table}

\section{Conclusions}
A novel surface decomposition method has been proposed for highly efficient and accurate sensitivity analysis of the first-passage dynamic reliability of linear systems subjected to Gaussian random excitations. To circumvent the direct computation of a complex integral over a non-smooth system limit-state hypersurface involved in the reliability sensitivity analysis, the proposed method decomposes this complex surface integral into a sum of simpler surface integrals over the constrained component hyperplanes. These component surface integrals can be evaluated at high efficiency for two main reasons: i) closed-form expressions are available for the component limit-state functions and their sensitivities for a linear system, which results in an explicit simulation of the component surface integral; ii) each constrained component hyperplane lies entirely within the mutual safe domain of all the other component limit-state functions, which leads to a straightforward generation of samples on the constrained component hyperplane. An effective importance sampling strategy is introduced to estimate the sum of all these component surface integrals, which is essential when a large number of component limit-state functions are involved. 

The effectiveness of the surface decomposition method has been validated by three first-passage dynamic reliability problems of linear systems, one of which involves a large-scale building frame structure. It has been found that, for the numerical examples studied, only less than 2000 function evaluations are required to obtain the sensitivity estimates of the first-passage failure probability with respect to all the design parameters with a target coefficient of variation of 0.1. The surface decomposition method exhibits a higher efficiency than the directional importance sampling with a speed-up ratio ranging from 1.29 to 2.56. Furthermore, similar to the efficient importance sampling and the domain decomposition method for first-passage dynamic reliability analysis, the surface decomposition method for reliability sensitivity analysis also demonstrates a higher efficiency as the failure probability becomes smaller. Numerical studies also indicate that the efficiency of the proposed method is not affected by the input dimensionality, while it decreases to a certain degree as the number of component failure events increases.

The core contribution of this work is that it represents the first attempt to explicitly address the complex surface integral arising in reliability sensitivity analysis, which has long been regarded as highly challenging
and thus largely bypassed in the literature. By introducing a novel yet simple concept of surface decomposition, the proposed method allows efficient evaluation of this integral for linear systems under Gaussian excitations and exhibits favorable performance compared with existing state-of-the-art approaches, while relying on an algorithm that is straightforward to implement and readily reproducible. A key advantage of the proposed method is its ability to reuse the function evaluation results for sensitivity analyses with respect to different design parameters, thereby substantially reducing the computational cost when a large number of parameters are considered. This feature highlights the great potential of the method for reliability-based parametric and topology optimization.

The current surface decomposition method has two main limitations: system linearity and excitation Gaussianity. The root cause of these limitations is that the efficient implementation of the method relies on the availability of closed-form expressions for dynamic responses and response sensitivities at different time instants. Without the assumptions of linear system behavior and Gaussian excitation, such closed-form expressions are either unavailable or significantly more difficult to obtain. The first limitation can be mitigated by incorporating the equivalent linearization technique, although this approach is generally restricted to mildly nonlinear systems and inevitably entails some loss of accuracy in reliability sensitivity estimation. The second limitation can be addressed for certain types of non-Gaussian excitations with specific forms, such as the quadratic Gaussian excitations, for which the explicit expressions of dynamic responses and response sensitivities can still be derived in quadratic form. A method to address this limitation is currently under investigation by the authors.

\section{Acknowledgements}
This work was supported by the start-up fund and the seed fund for Basic Research for New Staff from the University of Hong Kong. The first author and the corresponding author gratefully acknowledge the support of the University of Hong Kong.

\bibliography{Ref}

\appendix
\section{Efficient importance sampling}\label{appendix a}
Importance sampling rewrites the system failure probability integral in Eq.\eqref{system failure probability} as follows:
\begin{equation}\label{IS}
    \mathbb{P}
    = \int_{\mathbb{R}^n} \mathcal{H}(-G(\vect x)) \frac{f_{\vect X}(\vect x)}{h_A(\vect x)}h_A(\vect x)\, d\vect x
\end{equation}
where $h_A(\vect x)$ denotes the importance sampling PDF, which should cover the entire system failure domain $\left\{ G(\vect X)  \leq0 \right\}$. 

A very efficient importance sampling PDF has been proposed in \cite{au2001first}, which is given by
\begin{equation}\label{ID by Au and Beck}
    h_A(\vect x)
    =\sum_{k=1}^{m}\sum_{i=1}^{n}w_{ki} h_{ki}^{\star}(\vect x)
\end{equation}
where 
\begin{equation}\label{optimal component ID}
    h_{ki}^{\star}(\vect x)= \frac{\mathcal{H}(-g_{ki}(\vect x)) f_{\vect X}(\vect x)}{\mathbb{P}_{ki}}\quad(i=1,2,\dots,n;k=1,2,\dots,m)
\end{equation}
is the theoretically optimal importance sampling PDF corresponding to the component limit-state function $g_{ki}(\vect x)$; and 
\begin{equation}\label{weight}
    w_{ki}= \frac{\mathbb{P}_{ki}}{\sum_{j=1}^{m}\sum_{s=1}^{n}\mathbb{P}_{js}}\quad(i=1,2,\dots,n;k=1,2,\dots,m)
\end{equation}
is the weight coefficient associated with the importance sampling PDF $h_{ki}^{\star}(\vect x)$.

Substitution of Eqs.\eqref{ID by Au and Beck}-\eqref{weight} into Eq.\eqref{IS} yields the following importance sampling estimator:
\begin{equation}\label{IS estimator}
\begin{aligned}
\mathbb{P} &= Z_A
\int_{\mathbb{R}^n} 
\frac{\mathcal{H}(-G(\vect x))}
{\sum_{k=1}^{m}\sum_{i=1}^{n}\mathcal{H}(-g_{ki}(\vect x))}
h_A(\vect x)\, d\vect x \\[6pt]
&\approx 
\frac{Z_A}{N_A}\sum_{j=1}^{N_A}
\frac{1}
{\sum_{k=1}^{m}\sum_{i=1}^{n}\mathcal{H}(-g_{ki}(\vect x_j))}
\end{aligned}
\end{equation}
where $Z_A= \sum_{j=1}^{m}\sum_{s=1}^{n}\mathbb{P}_{js}$ is a constant; and $\vect x _j\;(j=1,2,\dots,N_A)$ are samples generated from $h_A(\vect x)$ in Eq.\eqref{ID by Au and Beck}, with $N_A$ being the number of samples.

\section{Finite difference method based on importance sampling}\label{appendix b}
Finite difference method estimates the sensitivity of system failure probability as follows:
\begin{equation}\label{FDM}
    \frac{\partial\mathbb{P}}{\partial \theta}
    \approx \frac{\mathbb{P}(\theta+\Delta \theta)-\mathbb{P}(\theta)}{\Delta \theta}
\end{equation}
where $\Delta \theta$ denotes a small perturbation in the design parameter $\theta$; and $\mathbb{P}(\theta)$ and $\mathbb{P}(\theta+\Delta \theta)$ denote the system failure probabilities corresponding to $\theta$ and its perturbed value $\theta+\Delta \theta$, respectively.

The main task of Eq.\eqref{FDM} is to estimate the change of probability due to the small perturbation $\Delta \theta$, which can be expressed as
\begin{equation}\label{probability change}
    \mathbb{P}(\theta+\Delta \theta)-\mathbb{P}(\theta)= \int_{\mathbb{R}^n} \left [ \mathcal{H}(-G(\vect x;\theta+\Delta \theta))-\mathcal{H}(-G(\vect x;\theta))\right ]  f_{\vect X}(\vect x)\, d\vect x
\end{equation}
where $G(\vect x;\theta)$ and $G(\vect x;\theta+\Delta \theta)$ are the system limit-state functions corresponding to $\theta$ and its perturbed value $\theta+\Delta \theta$, respectively.

Importance sampling can also be used to estimate the probability integral in Eq.\eqref{probability change}. Similar to Eq.\eqref{IS}, Eq.\eqref{probability change} can be rewritten as 
\begin{equation}\label{probability change IS}
    \mathbb{P}(\theta+\Delta \theta)-\mathbb{P}(\theta)= \int_{\mathbb{R}^n} \left [ \mathcal{H}(-G(\vect x;\theta+\Delta \theta))-\mathcal{H}(-G(\vect x;\theta))\right ]  \frac{f_{\vect X}(\vect x)}{h_B(\vect x)}h_B(\vect x)\, d\vect x
\end{equation}
A valid importance sampling PDF $h_B(\vect x)$ should cover the incremental failure domain induced by the small perturbation $\Delta \theta$.

Motivated by the construction of importance sampling PDF $h_A(\vect x)$ shown in Eq.\eqref{ID by Au and Beck}, a valid importance sampling PDF $h_B(\vect x)$ for Eq.\eqref{probability change IS} can be constructed as
\begin{equation}\label{ID FDM}
    h_B(\vect x)
    =\sum_{k=1}^{m}\sum_{i=1}^{n}w_{ki}(\theta) h_{ki}^{\star}(\vect x;\theta)+\sum_{k=1}^{m}\sum_{i=1}^{n}w_{ki}(\theta+\Delta\theta) h_{ki}^{\star}(\vect x;\theta+\Delta\theta)
\end{equation}
where 
\begin{equation}\label{optimal component ID FDM1}
    h_{ki}^{\star}(\vect x;\theta)= \frac{\mathcal{H}(-g_{ki}(\vect x;\theta)) f_{\vect X}(\vect x)}{\mathbb{P}_{ki}(\theta)}\quad(i=1,2,\dots,n;k=1,2,\dots,m)
\end{equation}
and
\begin{equation}\label{optimal component ID FDM2}
    h_{ki}^{\star}(\vect x;\theta+\Delta\theta)= \frac{\mathcal{H}(-g_{ki}(\vect x;\theta+\Delta\theta)) f_{\vect X}(\vect x)}{\mathbb{P}_{ki}(\theta+\Delta\theta)}\quad(i=1,2,\dots,n;k=1,2,\dots,m)
\end{equation}
are the theoretically optimal importance sampling PDFs corresponding to the component limit-state functions $g_{ki}(\vect x;\theta)$ and $g_{ki}(\vect x;\theta+\Delta \theta)$, respectively; $\mathbb{P}_{ki}(\theta)$ and $\mathbb{P}_{ki}(\theta+\Delta \theta)$ are the corresponding component failure probabilities with analytical solutions available; and 
\begin{equation}\label{weight FDM1}
    w_{ki}(\theta)= \frac{\mathbb{P}_{ki}(\theta)}{\sum_{j=1}^{m}\sum_{s=1}^{n} \left [ \mathbb{P}_{js}(\theta)+\mathbb{P}_{js}(\theta+\Delta \theta)\right ]}\quad(i=1,2,\dots,n;k=1,2,\dots,m)
\end{equation}
and
\begin{equation}\label{weight FDM2}
    w_{ki}(\theta+\Delta \theta)= \frac{\mathbb{P}_{ki}(\theta+\Delta \theta)}{\sum_{j=1}^{m}\sum_{s=1}^{n}\left [ \mathbb{P}_{js}(\theta)+\mathbb{P}_{js}(\theta+\Delta \theta)\right ]}\quad(i=1,2,\dots,n;k=1,2,\dots,m)
\end{equation}
are the weight coefficients associated with the importance sampling PDFs $h_{ki}^{\star}(\vect x;\theta)$ and $h_{ki}^{\star}(\vect x;\theta+\Delta \theta)$, respectively.

Substituting Eqs.\eqref{ID FDM}-\eqref{weight FDM2} into Eq.\eqref{probability change IS}, one can obtain the importance sampling estimator as follows:
\begin{equation}\label{IS FDM estimator}
\begin{aligned}
\mathbb{P}(\theta+\Delta \theta)-\mathbb{P}(\theta) 
&= Z_B
\int_{\mathbb{R}^n} 
\frac{ \mathcal{H}(-G(\vect x;\theta+\Delta \theta))-\mathcal{H}(-G(\vect x;\theta))}
{\sum_{k=1}^{m}\sum_{i=1}^{n}\left [ \mathcal{H}(-g_{ki}(\vect x;\theta))+\mathcal{H}(-g_{ki}(\vect x;\theta+\Delta\theta))\right ] }
h_B(\vect x)\, d\vect x \\[6pt]
&\approx \frac{Z_B}{N_B}\sum_{j=1}^{N_B}
\frac{ \mathcal{H}(-G(\vect x_j;\theta+\Delta \theta))-\mathcal{H}(-G(\vect x_j;\theta)) }
{\sum_{k=1}^{m}\sum_{i=1}^{n}\left [\mathcal{H}(-g_{ki}(\vect x_j;\theta))+\mathcal{H}(-g_{ki}(\vect x_j;\theta+\Delta \theta))\right ]}
\end{aligned}
\end{equation}
where $Z_B= \sum_{j=1}^{m}\sum_{s=1}^{n}\left [\mathbb{P}_{js}(\theta)+\mathbb{P}_{js}(\theta+\Delta \theta)\right]$ is a constant; and $\vect x _j\;(j=1,2,\dots,N_B)$ are samples generated from $h_B(\vect x)$ in Eq.\eqref{ID FDM}, with $N_B$ being the number of samples.

Note that the support of the importance sampling PDF $h_B(\vect x)$ is exactly the union of the two failure domains, $\left\{ G(\vect X;\theta)  \leq0 \right\}$ and $\left\{ G(\vect X;\theta+\Delta \theta)  \leq0 \right\}$, which sufficiently covers the incremental failure domain induced by the small perturbation $\Delta \theta$. Although the sampling region has been significantly concentrated compared with the original sampling PDF $f_{\vect X}(\vect x)$, the efficiency of  $h_B(\vect x)$ remains limited, since the target sampling region is typically very small depending on the magnitude of the perturbation $\Delta \theta$. Nevertheless, the present finite difference method in conjunction with importance sampling provides a theoretically correct and practically feasible approach for evaluating the sensitivity of the failure probability, whereas using the original sampling PDF would be nearly impractical.

 \end{document}